\documentclass[a4paper,11pt]{article}
\pdfoutput=1
\usepackage{jheppub}
\usepackage[english]{babel}
\usepackage{amsthm,amsfonts}
\usepackage{accents}
\usepackage{dutchcal}
\usepackage{upgreek}
\allowdisplaybreaks
\usepackage{enumitem}
\usepackage{tikz}
\usetikzlibrary{arrows}
\usetikzlibrary{calc}
\tikzset{>=latex}
\usepackage{dsfont}
\usepackage{cite}
\usepackage{mcite}

\usepackage{todonotes}

\newcommand{\dd}{\mathrm{d}}

\newcommand{\PS}{Pol\'a\v{c}ek-Siegel}

\newcommand{\hrangle}{\raisebox{-3px}{\resizebox{6px}{12px}{$\succ$}}}
\newcommand{\hlangle}{\raisebox{-3px}{\resizebox{6px}{12px}{$\prec$}}}
\newcommand{\heta}{\upeta}
\newcommand{\htau}{\uptau}
\newcommand{\hatt}{\widehat{t}}

\newcommand{\tf}{\widetilde{f}}
\newcommand{\tkappa}{\widetilde{\kappa}}
\newcommand{\tA}{\widetilde{A}}
\newcommand{\txi}{\widetilde{\xi}}
\newcommand{\dxi}{\upxi}
\newcommand{\tSinv}{{\widetilde{S}}^{-1}}

\newcommand{\Hc}{\mathcal{H}}
\newcommand{\Lc}{\mathcal{L}}
\newcommand{\Rc}{\mathcal{R}}
\newcommand{\Pb}{\overline{P}}

\newcommand{\Al}{\overline{A}}
\newcommand{\Ar}{\underline{A}}
\newcommand{\Bl}{\overline{B}}

\newcommand{\Cl}{\overline{C}}
\newcommand{\Cr}{\underline{C}}
\newcommand{\Dl}{\overline{D}}
\newcommand{\Dr}{\underline{D}}

\newcommand{\Odd}[1][d]{\mathrm{O}(#1,#1)}
\newcommand{\GL}[1][n]{\mathrm{GL}(#1)}
\newcommand{\GS}{G_{\mathrm{S}}}
\newcommand{\GSL}{\underline{\GS}}
\newcommand{\GSR}{\overline{\GS}}
\newcommand{\gs}{\mathfrak{g}_{\mathrm{S}}}
\newcommand{\GD}{G_{\mathrm{D}}}
\newcommand{\GM}{G_{\mathrm{M}}}
\newcommand{\GPS}{G_{\mathrm{PS}}}
\newcommand{\gPS}{\mathfrak{g}_{\mathrm{PS}}}
\newcommand{\gS}{\mathfrak{g}_{\mathrm{S}}}

\newcommand{\Ah}{\hat{A}}
\newcommand{\Bh}{\hat{B}}
\newcommand{\Ch}{\hat{C}}
\newcommand{\AhL}{\overline{\hat{A}}}
\newcommand{\BhL}{\overline{\hat{B}}}
\newcommand{\ChL}{\overline{\hat{C}}}

\newcommand{\AhR}{\underline{\hat{A}}}
\newcommand{\BhR}{\underline{\hat{B}}}
\newcommand{\ChR}{\underline{\hat{C}}}

\newcommand{\Ac}{\mathcal{A}}
\newcommand{\Bc}{\mathcal{B}}
\newcommand{\Cc}{\mathcal{C}}
\newcommand{\Dc}{\mathcal{D}}

\newcommand{\genLieHet}{\mathcal{L}}

\newcommand{\ZHet}{\mathcal{Z}}

\newcommand{\cE}{\mathcal{E}}
\newcommand{\cA}{\mathcal{A}}
\newcommand{\cD}{\mathcal{D}}
\newcommand{\cF}{\mathcal{F}}
\newcommand{\cG}{\mathcal{G}}
\newcommand{\bcF}{\boldsymbol{\cF}}
\newcommand{\bcG}{\boldsymbol{\cG}}
\newcommand{\tbcG}{\boldsymbol{\widetilde{\cG}}}
\newcommand{\bF}{\boldsymbol{F}}

\newcommand{\THet}{\mathcal{T}}

\newcommand{\alphai}{\tilde{\alpha}}
\newcommand{\alphan}{\underaccent{\tilde}{\alpha}}
\newcommand{\alphaL}{\underline{\alpha}}
\newcommand{\alphaR}{\overline{\alpha}}
\newcommand{\betaL}{\underline{\beta}}
\newcommand{\betaR}{\overline{\beta}}
\newcommand{\gammaL}{\underline{\gamma}}
\newcommand{\gammaR}{\overline{\gamma}}
\newcommand{\deltaL}{\underline{\delta}}
\newcommand{\deltaR}{\overline{\delta}}
\newcommand{\epsilonL}{\underline{\epsilon}}

\newcommand{\rhoL}{\underline{\rho}}

\newcommand{\chiL}{\underline{\chi}}
\newcommand{\chiR}{\overline{\chi}}
\newcommand{\muL}{\underline{\mu}}

\newcommand{\nuL}{\underline{\nu}}

\newcommand{\sigmaL}{\underline{\sigma}}

\newcommand{\aL}{\underline{a}}
\newcommand{\aR}{\overline{a}}
\newcommand{\bL}{\underline{b}}
\newcommand{\bR}{\overline{b}}
\newcommand{\cL}{\underline{c}}
\newcommand{\cR}{\overline{c}}
\newcommand{\dL}{\underline{d}}
\newcommand{\dR}{\overline{d}}
\newcommand{\eL}{\underline{e}}
\newcommand{\eR}{\overline{e}}
\newcommand{\fL}{\underline{f}}
\newcommand{\fR}{\overline{f}}
\newcommand{\gL}{\underline{g}}
\newcommand{\gR}{\overline{g}}

\newcommand{\AcL}{\underline{\Ac}}
\newcommand{\AcR}{\overline{\Ac}}
\newcommand{\BcL}{\underline{\Bc}}
\newcommand{\BcR}{\overline{\Bc}}
\newcommand{\CcL}{\underline{\Cc}}
\newcommand{\CcR}{\overline{\Cc}}

\newcommand{\llangle}{\langle\!\langle}
\newcommand{\rrangle}{\rangle\!\rangle}

\newcommand{\cX}{\mathcal{X}}

\newcommand{\Khet}{\mathcal{K}}

\newsavebox\MBox

\title{\boldmath Unraveling the generalized\\Bergshoeff-de Roo identification}

\author[a]{Achilleas Gitsis}
\author[a]{and Falk Hassler}

\emailAdd{achilleas.gitsis@uwr.edu.pl}
\emailAdd{falk.hassler@uwr.edu.pl}

\affiliation[a]{University of Wrocław, Faculty of Physics and Astronomy, Maksa Borna 9, 50-204 Wrocław, Poland}

\abstract{We revisit duality-covariant higher-derivative corrections which arise from the generalized Bergshoeff-de Roo (gBdR) identification, a prescription that gives rise to a two parameter family of $\alpha'$-corrections to the low-energy effective action of the bosonic and the heterotic string. Although it is able to reproduce all corrections at the leading and sub-leading ($\alpha'^2$) order purely from symmetry considerations, a geometric interpretation, like for the two-derivative action and its gauge transformation is lacking. To address this issue and to pave the way for the future exploration of higher-derivative (=higher-loop for the $\beta$-functions of the underlying $\sigma$-model) corrections to generalized dualities, consistent truncations and integrable $\sigma$-models, we recover the gBdR identification's results from the \PS{} construction that provides a natural notion of torsion and curvature in generalized geometry.}

\begin{document}

\maketitle

\section{Introduction}
Gravity is perturbatively non-renormalizable and should therefore only be seen as an effective theory that requires modification when approaching high energies in the ultraviolet (UV) regime. Currently, it is still debated what its correct UV completion is and while there are candidates, like string theory, even without a complete knowledge of the fundamental theory, consistency conditions can be leveraged to rule out theories that are incompatible with a putative theory of quantum gravity.

A common feature of different approaches to quantizing gravity is the emergence of new symmetries. They have to be compatible with all directly observed symmetries, like reparameterizations (diffeomorphisms) and the gauge transformations that accompany additional degrees of freedom interacting with gravity. Further symmetry transformations that go beyond this obvious part will restrict admissible quantum corrections. Here, the guiding principle is: The more restrictive, the better (as long as no observations are contradicted), because it increases the predictive power of the assumed symmetry. As an example of this approach consider the two-derivative action
\begin{equation}\label{eqn:SNSNS}
	S = \int \dd^{d} x\, \sqrt{g}\, e^{-2\phi}\left( R + 4(\partial\phi)^2 - \frac{1}{12} H^2 \right) \qquad \text{with} \qquad H = \dd B\,.
\end{equation}
It is a part of the low-energy effective actions of the five perturbative superstring theories ($d=10$) and the bosonic string ($d=26$). Diffeomorphisms, mediated by the Lie derivative, and $B$-field gauge transformations, $B \rightarrow B + \dd A$, leave \eqref{eqn:SNSNS} invariant. Remarkably, these two symmetries can be embedded into the larger group $\mathrm{O}(d,d)$ by introducing the generalized metric
\begin{equation}
	\Hc_{IJ} = \begin{pmatrix} g_{ij} - B_{ik}g^{kl}B_{lj} & B_{ik}g^{kj} \\ - g^{ik}B_{kj} & g^{ij} \end{pmatrix}\,,
\end{equation}
which unifies the metric $g_{ij}$ and the $B$-field $B_{ij}$ into a single object, the $\mathrm{O}(d,d)$-invariant metric\footnote{It is used in combination with its inverse $\eta^{IJ}$ to raise/lower capital Latin indices.}
\begin{equation}
	\eta_{IJ} = \begin{pmatrix} 0 & \delta_i^j \\ \delta_j^i & 0 \end{pmatrix}\,,
\end{equation}
and the generalized Lie derivative \cite{Hohm:2010pp}
\begin{equation}\label{eqn:genLieGenMetric}
	\delta_\xi \Hc_{IJ} = \Lc_\xi \Hc_{IJ} = \xi^K \partial_K \Hc_{IJ} - 2 \left( \partial_K \xi_{(I} - \partial_{(I} \xi_K \right) \Hc_{J)}{}^K \,.
\end{equation}
In contrast to the standard Lie derivative, the latter only closes after imposing the strong constraint $\partial_I \partial^I \,\cdot\,= 0$ for all fields, parameters of transformations like $\xi^I$ and products of them. If not otherwise stated, we implement it by requiring $\partial_I \, \cdot \, = \begin{pmatrix} \partial_i \, \cdot \, & 0 \end{pmatrix}$. At this point, it is important to note that the group $\mathrm{O}(d,d)$ contains $d(d-1)/2$ more generators than required for diffeomorphisms and $B$-field transformations. Hence, it is more restrictive and capable of fixing the relative factors between the three terms in the action \eqref{eqn:SNSNS}. The same feat can be achieved by discussing supersymmetry, which we will not do here. It is possible to rewrite the action such that its full symmetry is manifest by introducing the generalized Ricci scalar and dilaton as \cite{Hohm:2010pp}
\begin{equation}\label{eqn:S1}
	S = \int \dd^d x \, e^{-2 \Phi} \, \Rc
	\qquad \text{with} \qquad
	\Phi = \phi - \frac12 \log\det(g_{ij})\,.
\end{equation}
As hinted by the name, $\Rc$ admits an interpretation as the Ricci scalar of a generalized connection, which is a consequence of the $\mathrm{O}(d,d)$ symmetry. It is discussed in the framework of double field theory \cite{Siegel:1993xq,*Siegel:1993th,*Hull:2009mi} and the closely related generalized geometry \cite{Coimbra:2011nw}. In contrast to standard geometry, there is however the problem that an analog of the fundamental theorem of Riemannian geometry, where metricity and vanishing torsion are sufficient to completely fix the connection, has not been established yet. Therefore, generalized connections have to cope with undetermined components \cite{Hohm:2011si}.

Considering higher-derivative corrections, as they are inevitable for an effective action like \eqref{eqn:SNSNS}, the situation becomes more opaque. In this case, generalized diffeomorphisms, as they are mediated by \eqref{eqn:genLieGenMetric}, alone are not enough \cite{Hohm:2016lge} to reproduce the corrections predicted by string theory. Additionally, double Lorentz transformations have to be taken into account. They are motivated by the observation that the generalized metric is not an unconstrained element of $\mathrm{O}(d,d)$ but rather valued in the coset
\begin{equation}
	\Hc_{IJ} \in \frac{\mathrm{O}(d,d)}{\mathrm{O}(1,d-1) \times \mathrm{O}(d-1,1)},
\end{equation}
where the denominator is called double Lorentz group. To make this statement more explicit, we express the generalized metric in terms of a generalized frame $E^A{}_I$, which is an unconstrained element of $\mathrm{O}(d,d)$, as
\begin{equation}
	\Hc_{IJ} = E^A{}_I E^B{}_J \Hc_{AB}\, ,
	\quad \text{while additionally imposing} \quad
	\eta_{IJ} = E^A{}_I E^B{}_J \eta_{AB}\,.
\end{equation}
In this equation both $\Hc_{AB}$ and $\eta_{AB}$ are constant and invariant under double Lorentz transformations. Consequentially, one might now use either the generalized metric and \eqref{eqn:genLieGenMetric} or the corresponding frame with
\begin{equation}\label{eqn:deltaFrame}
	\delta_{(\xi,\Lambda)} E^A{}_I = \Lc_\xi E^A{}_I + \Lambda^A{}_B E^B{}_I
\end{equation}
to capture the symmetries of \eqref{eqn:SNSNS}. The new parameter $\Lambda_{AB} = -\Lambda_{BA}$ that appears here generates the double Lorentz group. In the following it will be necessary to project separately one each of its two factors. This is done with the two projectors
\begin{equation}
	P_{AB} = \frac12 \left( \eta_{AB} + \Hc_{AB} \right)
	\quad \text{and} \quad
	\Pb_{AB} = \frac12 \left( \eta_{AB} - \Hc_{AB} \right)\,,
\end{equation}
respectively. Writing them for every index is cumbersome, especially in longer relations. Therefore, the shorthand notation $V_{\Al} = P_A{}^B V_B$ and $V_{\Ar} = \Pb_A{}^B V_B$ is commonly used.

A crucial observation in capturing higher-derivative corrections has been: While it is possible to maintain the form of the generalized Lie derivative in \eqref{eqn:deltaFrame}, double Lorentz transformations must be modified to \cite{Marques:2015vua}
\begin{equation}
	\delta E_{\Ar}{}^I E_{\Bl I} = \frac{a}2 D_{\Ar} \Lambda^{\Cr\Dr} F_{\Bl\Cr\Dr} + \frac{b}2 D_{\Bl} \Lambda^{\Cl\Dl} F_{\Ar\Cl\Dl}\, ,
\end{equation}
where
\begin{equation}
	F_{ABC} = 3 D_{[A} E_B{}^I E_{C]I}
\end{equation}
are called generalized fluxes and
\begin{equation}
	D_A = E_A{}^I \partial_I
\end{equation}
is referred to as flat derivative. Note that two parameters, $a$ and $b$, control the deformation of the double Lorentz symmetry -- one for each factor. There is strong physical support for this deformation by the Green-Schwarz (GS) anomaly cancellation mechanism \cite{Green:1984sg} of the heterotic string. Thus, they are called generalized GS (gGS) transformations. With all symmetries fixed, it is again possible to construct a unique invariant action \cite{Marques:2015vua}
\begin{equation}
	S = \int \dd^d x \, e^{-2\Phi} \left( \Rc + a \Rc^{(-)} + b \Rc^{(+)} \right),
\end{equation}
up to field redefinitions. It is inspired by earlier work on four-derivative corrections for the heterotic string \cite{Bergshoeff:1989de}, now known as Bergshoeff-de Roo (BdR) identification and provides the correct, leading order $\alpha'$-corrections for heterotic ($a=-\alpha', b=0$), bosonic ($a=b=-\alpha'$) and type II (super)strings ($a=b=0$). Committing to a non-vanishing value for either $a$ or $b$ requires an infinite tower of higher-derivative corrections, which can be obtained by a technique dubbed generalized Bergshoeff-de Roo (gBdR) identification \cite{Baron:2018lve,Baron:2020xel}. Beyond the obvious appeal of better understanding the structure of admissible corrections to the leading order action \eqref{eqn:SNSNS}, the gBdR identification has proven extremely useful in the context of generalized dualities \cite{Hassler:2020tvz,Borsato:2020wwk,Codina:2020yma,Hassler:2020wnp} and consistent truncations \cite{Baron:2017dvb}. At the same time, however, it comes with the challenges that
\begin{enumerate}[label=C\arabic*)]
	\item\label{db:noGeom} Despite their suggestive name, both $\Rc^{(-)}$ and $\Rc^{(+)}$ lack the geometric interpretation of the leading order contribution $\Rc$ as the Ricci scalar in generalized geometry. This is not just a conceptual imperfection because recently it became clear that important applications of generalized geometry and extended field theories, like the above mentioned generalized dualities and consistent truncations, heavily benefit from a geometric understanding \cite{Butter:2022iza,Hassler:2023axp}.
\end{enumerate}
The situation is even more problematic for corrections beyond the leading order in the gBdR identification. While it successfully reproduces known corrections, it comes with two additional questions, namely
\begin{enumerate}[resume*]
	\item\label{db:strangeMath} The mathematical structure underlying the gBdR identification is not completely clear. It requires sending the dimension of a semi-simple Lie algebra to infinity requiring a cumbersome regularization process to obtain corrections beyond the leading order. Therefore, the authors of \cite{Baron:2018lve} conclude that it should be understood as a prescription whose fundamental principles require further study. Similar to~\ref{db:noGeom}, this has at least one important practical implication: With increasing number of derivatives, the obtained results become extremely complicated quickly. But the original gBdR identification does not permit to simplify intermediate results; it is all or nothing. Thus, a way to break down the computation into more manageable parts is desirable, and most likely even inevitable, in order to learn more about the structure of higher-derivative corrections in string theory.
	\item\label{db:zeta3} It has been argued that by deforming double Lorentz transformations as done in the gBdR construction, it is not possible to introduce new deformation parameters beyond $a$ and $b$. This becomes a severe problem when reaching eight derivatives because at this order, known corrections from string theory are weighted by $\zeta(3)$, a transcendental number which cannot arise from rational functions of $a$ and $b$ \cite{Hronek:2020xxi}. As all these results have been obtained from a bottom-up perspective and they do not explain how this obstruction arises from a geometrical point of view.
\end{enumerate}

This article aims to make progress on~\ref{db:noGeom} and~\ref{db:strangeMath} by using a tool called \PS{} construction \cite{Polacek:2013nla}. As some of us have shown in earlier work, it provides a tool to obtain covariant curvature and torsion tensors in generalized geometry and double field theory \cite{Butter:2022iza}. Covariance is with respect to generalized diffeomorphisms and an additional symmetry group $\GS$, that is usually chosen to be either the double Lorentz group or a subgroup of it. This approach has two major strengths: It naturally generates covariant torsion and curvature tensors by unifying all generalized connections and the generalized frame into an element of the Lie group $\GPS$. Note the use of the plural, namely connections, here. That is because generalized geometry features a gauge-for-gauge symmetry, a hierarchy of connections \cite{Hassler:2023axp} arises. They are organized according to the tensor hierarchy \cite{deWit:2008ta} known from gauged supergravity. Furthermore, the \PS{} construction allows identifying the origin of all known generalized dualities and the related consistent truncations as generalized cosets, the lift of homogeneous spaces to generalized geometry \cite{Butter:2022iza}. Building on these tools, we show here how the gBdR identification arises in the heterotic version of the \PS{} construction (introduced recently in \cite{HasslerHetPS:2024}) through torsion constraints and a partial gauge fixing. A central step in this effort is to identify the generalized structure group $\GS$. It is closely related to the infinite simple group in the original identification but we obtain it recursively from the double Lorentz group -- thereby revealing its hierarchical structure. For most computations it is sufficient to work only with a finite number of $\GS$'s generators. In particular, this allows to obtain intermediate results which can be used in the computation of higher orders. Only in the final step one has to deal with a particular twist of generators that arises in the identification and requires an infinite tower of generators. Summing over this tower requires a regularization, which has to be compatible with the residual symmetry after the partial gauge fixing we perform. At this point, it is important to mention that the final results we obtain are the same as for the original identification in \cite{Baron:2020xel}. We focus mostly on generalized GS transformations because it is easier to compare them. But, as we will argue, the resulting invariant actions also match. The main selling point for our approach is that it makes more of the underlying structure visible. For example, we are able to present a simple, universal expression for the generalized GS transformations with up to four derivatives which we conjecture to hold for all orders.

Challenge~\ref{db:zeta3}, is not addressed here. But at a first glance, there is considerable freedom in fixing $\GPS$ and $\GS$. While we choose them to recover the results from the gBdR identification, it is conceivable to refine this choice to make more deformation parameters accessible. The present article consists of two parts. In section~\ref{sec:tPS}, we review the \PS{} construction for heterotic strings presented in \cite{HasslerHetPS:2024} and restrict it to a non-degenerate quadratic form $\kappa_{\alpha\beta}$ as the relevant setup for the gBdR identification. Afterwards, we proceed in section~\ref{sec:gBdRi} with the construction of $\GS$ and find a systematic way to organize its generators. In this process, we will encounter two different bases for them which are relevant to our discussion. Although both are related by a similarity transformation, a single generator in one of them will turn into a full tower of generators in the other. In section~\ref{sec:collapsingTowers}, we thus present a technique -- dubbed collapsing towers -- to deal with the resulting large but mostly redundant towers of generators. Equipped with all the necessary tools, sections~\ref{sec:identification} and~\ref{sec:genGSTr} eventually present the torsion constraints and gauge fixing procedure which form the core of our approach. They will recover the generalized Green-Schwarz (gGS) transformation up to four derivatives presented in \cite{Baron:2020xel} while the corresponding action and its correction are discussed in section~\ref{sec:PSaction}. Finally, we conclude with section~\ref{sec:conclusions}.

\section{Twisted \PS{} construction}\label{sec:tPS}
Riemann curvature and torsion are central concepts in differential geometry, but their extension to generalized geometry is not straightforward. A generalized version of the torsion tensor can be defined based on the generalized Lie derivative, but the generalized Riemann tensor is more elusive. A first naive guess -- the Riemann tensor of standard differential geometry -- does not transform covariantly under generalized diffeomorphisms and double Lorentz transformations. It has to be further modified and projected. Pol\'a\v{c}ek and Siegel \cite{Polacek:2013nla} proposed to resort to an extended version of the physical space-time $M$ with a flat generalized connection\footnote{A flat connection has only torsion but no curvature.} to overcome this problem. We will call this extended space mega-space to hint that it is even larger than the usual doubled space of $M$ used in double field theory. Especially recently, this approach has proven very useful in exploring generalized dualities, consistent truncations and the geometry of extended field theories.  A still open question is how higher-derivative corrections fit in this picture. To answer it, we will consider a twisted version \cite{} of the original construction, which is capable of describing gauged double field theory as it arises for the heterotic string. Considering that we eventually want to make contact with the gBdR identification, this is the natural starting point. In the following, we will review its most important aspects.

\subsection{Algebra}\label{sec:algebra}
From an abstract point of view, the \PS{} construction takes some algebraic data as input to produce covariant tensors as output. Saying that, once the algebraic part is understood, everything else follows the algorithm. To understand the gBdR identification in this framework, we take the duality group $\GD$ of the physical space as
\begin{equation}
	\GD = \Odd[d]\,.
\end{equation}
It is embedded in the duality group of the mega-space, $\GM=\Odd[n+d]$. The structure of this group is very important for the inner workings of the \PS{} construction, which have been worked out in \cite{}. Here, we do not have to deal with them. Instead, we only need the subgroup $\GPS\subset\GM$, which is generated by
\begin{equation}\label{eqn:KABandRs}
	\gPS = \{ K_{AB}\,, R^A_\alpha\,, R_{\alpha\beta} \}\,.
\end{equation}
The generator $K_{AB}$ generates $\GD$ through
\begin{equation}\label{eqn:commKABKCD}
	[ K_{AB}, K_{CD} ]  = - \tfrac12 ( \eta_{AC} K_{BD} - \eta_{AD} K_{BC} + \eta_{BD} K_{AC} - \eta_{BC} K_{AD} ) = 2 \eta_{[A|[C} K_{D]|B]}\,,
\end{equation}
with indices $A=1,\ldots,2 d$ and where $\eta_{AB}$ denotes the invariant metric of $\Odd[d]$. All remaining generators are governed by the non-vanishing commutators ($\alpha=1,\ldots,n$)
\begin{align}\label{eqn:commR1R1}
	[ R^A_\alpha, R^B_\beta ]             & = \eta^{AB} R_{\alpha\beta} - 2 K^{AB} \kappa_{\alpha\beta}\,, \\\label{eqn:commR1K}
	[ K_{AB}, R^C_\gamma ]                & = - \delta_{[A}^C \eta_{B]D} R^D_\gamma\,,                     \\
	\label{eqn:commR1R2}
	[ R^A_\alpha, R_{\beta\gamma} ]       & = - 2 \kappa_{\alpha[\beta} R_{\gamma]}^A\,, \qquad \text{and} \\\label{eqn:commR2R2}
	[ R_{\alpha\beta}, R_{\gamma\delta} ] & = - 4 \kappa_{[\alpha| [\gamma} R_{\delta]|\beta]}\,.
\end{align}
We assume that the tensor $\kappa_{\alpha\beta}$ that appears here is symmetric under the exchange of its two indices. Furthermore, it will be non-degenerate with the inverse $\kappa^{\alpha\beta}$. As $\eta_{AB}$/$\eta^{AB}$ is used to lower/raise capital Latin indices, we use $\kappa_{\alpha\beta}$/$\kappa^{\alpha\beta}$ to do the same with Greek indices. This is a special case of the more general pairings analyzed in \cite{HasslerHetPS:2024} whose refined indices $\alpha = \begin{pmatrix} \alphai & \alphan \end{pmatrix}$ allow for null-directions, implying $\kappa_{\alphan\beta}=0$. As there are no null-directions here, we can safely remove all $\alphan$ indices and at the same time remove the tilde from $\alphai$ to avoid cluttering the notation. Finally, we will also need the $\GL[n]$-generators
\begin{equation}
	[ K_\alpha^\beta, K_\gamma^\delta ] = \delta_\alpha^\delta K_\gamma^\beta - \delta_\gamma^\beta K_\alpha^\delta
\end{equation}
which will be used in section~\ref{sec:GS} to describe the generalized structure group $\GS$.

After the algebra is fixed, the next step is to find its fundamental representation. A detailed overview of how to do this can be found in \cite{HasslerHetPS:2024}. Here, we focus on the ``heterotic basic'' formed by the dual basis vectors
\begin{equation}\label{eqn:decompAh}
	|^{\Ac}\hrangle = \begin{pmatrix} |^A\hrangle \quad & |_\alpha\hrangle \end{pmatrix}\,, \qquad \text{and} \qquad
	\hlangle_{\Ac}| = \begin{pmatrix} \langle_A | \quad & \hlangle^\alpha | \end{pmatrix}{}^T
\end{equation}
that satisfy
\begin{equation}
	\hlangle_{\Ac}|^{\Bc} \hrangle = \delta_{\Ac}^{\Bc}\,.
\end{equation}
Following \cite{HasslerHetPS:2024}, one computes the matrices representing the generators of $\GPS$ because only these are relevant for the present work. We start with
\begin{align}\label{eqn:matrixKAB}
	\hlangle_{\Ac}| K_{CD} |^{\Bc}\hrangle           & =  \begin{pmatrix}
		                                                      \eta_{A[C} \delta_{D]}^B & 0 \\
		                                                      0                        & 0
	                                                      \end{pmatrix}                                                 \,,
	\intertext{followed by}
	\hlangle_{\Ac}| R_\gamma^C |^{\Bc}\hrangle       & = \begin{pmatrix}
		                                                     0                               & \kappa_{\beta\gamma} \delta_A^C \\
		                                                     -\delta_\gamma^\alpha \eta^{BC} & 0
	                                                     \end{pmatrix}\,, \qquad\text{and}  \\[0.2em]
	\hlangle_{\Ac}| R_{\gamma\delta} |^{\Bc}\hrangle & = \begin{pmatrix}
		                                                     0 & 0                                               \\
		                                                     0 & 2 \kappa_{\beta[\gamma} \delta_{\delta]}^\alpha
	                                                     \end{pmatrix}\,.
\end{align}
In particular, one finds that all of them leave the metric
\begin{equation}
	\heta_{\Ac\Bc} = 	\begin{pmatrix}
		\eta_{AB} & 0                    \\
		0         & \kappa^{\alpha\beta}
	\end{pmatrix}
\end{equation}
of heterotic double field theory invariant. Assuming further that $\kappa_{\alpha\beta}$ has signature $(p,q)$, we see that
\begin{equation}\label{eqn:GPS}
	\GPS = \mathrm{O}(d+p, d+q)\,.
\end{equation}
Accordingly, we combine its generators to
\begin{equation}\label{eqn:Khet}
	\Khet_{\Ac\Bc} = \begin{pmatrix}
		K_{AB} \quad              & - \tfrac12 R_A^\beta       \\
		\tfrac12 R^\alpha_B \quad & - \tfrac12 R^{\alpha\beta}
	\end{pmatrix}\,,
\end{equation}
governed by the commutator
\begin{equation}\label{eqn:commKhetKhet}
	[ \Khet_{\Ac\Bc}, \Khet_{\Cc\Dc} ] = 2 \heta_{[\Ac|[\Cc} \Khet_{\Dc]|\Bc]} \,.
\end{equation}

\subsubsection{Generalized structure group}\label{sec:GS}
Another central object of the \PS{} construction is the generalized structure group $\GS$. Initially, we just assumed that it should be embedded into $\GL[n]$ with $n=p+q=\dim\GS$. To make further progress, we additionally assume that the Lie algebra generating $\GS$ is spanned by the generators $\hatt_\alpha$, satisfying the commutators
\begin{equation}\label{eqn:commtalphabeta}
	[ \hatt_\alpha, \hatt_\beta ] = - f_{\alpha\beta}{}^\gamma \hatt_\gamma\,.
\end{equation}
As explained in \cite{HasslerHetPS:2024}, its action has to leave $\kappa_{\alpha\beta}$ invariant, saying that
\begin{equation}\label{eqn:kappainvaraint}
	f_{\alpha(\beta}{}^\delta \kappa_{\gamma)\delta} = 0
\end{equation}
has to hold. Following further the argument there, we seek for the most general embedding with a non-trivial action on the subgroup generated by $\gPS$. This is achieved by decomposing
\begin{align}\label{eqn:deftalpha}
	\hatt_\alpha      & = K_\alpha + t_\alpha \qquad \text{with} \qquad K_{\alpha} = f_{\alpha\beta}{}^\gamma K_\gamma^\beta \,, \qquad \text{and}                                \\\label{eqn:talphadecomp}
	\gPS \ni t_\alpha & = (t_\alpha)_{\Bc\Cc} \Khet^{\Bc\Cc} = (t_\alpha)_{AB} K^{AB} + (t_\alpha)_B{}^{\gamma} R_\gamma^B + \tfrac12 (t_\alpha)^{\beta\gamma} R_{\gamma\beta}\,,
\end{align}
where $(t_\alpha)_{AB}$, $(t_\alpha)_B{}^\gamma$ and $(t_\alpha)^{\beta\gamma}$ are constants controlling the embedding. They can be conveniently combined into $(t_{\alpha})_{\Bc\Cc}$ which is antisymmetric with respect to the last two indices. However, it is not possible to choose them freely because \eqref{eqn:commtalphabeta} requires
\begin{equation}
	[ t_\alpha, t_\beta ] = - [ K_\alpha, t_\beta ] - [ t_\alpha, K_\beta ] - f_{\alpha\beta}{}^{\gamma} t_\gamma\,.
\end{equation}
Once the constants $(t_{\alpha})_{\Bc\Cc}$ are fixed, which we will do in subsection~\ref{sec:constrGS}, this equation has to be checked carefully. For now, we just want to point out that there is one more generator, namely
\begin{equation}
	R_\alpha = \tfrac12 f_{\alpha\beta}{}^\gamma \kappa^{\beta\delta} R_{\gamma\delta}\,.
\end{equation}
which can be understood as the natural counter part of $K_\alpha$ in $\gPS$. This can be seen from the commutators
\begin{align}
	[K_\alpha, K_\beta]         & = - f_{\alpha\beta}{}^\gamma K_\gamma\,, \qquad           &
	[R_\alpha, R_\beta]         & = - f_{\alpha\beta}{}^\gamma R_\gamma\,,
	\intertext{and the action on $\gPS$}
	[K_\alpha, K_{BC}]          & =0 \,, \qquad                                             & [R_\alpha, K_{BC}] & =0 \,, \\
	[K_\alpha, R_\beta^B]       & = - f_{\alpha\beta}{}^\gamma R_\gamma^B\,, \qquad         &
	[R_\alpha, R_\beta^B]       & = - f_{\alpha\beta}{}^\gamma R_\gamma^B\,,                                              \\
	[K_\alpha, R_{\beta\gamma}] & = 2 f_{\alpha[\beta}{}^\delta R_{\gamma]\delta}\,, \qquad &
	[R_\alpha, R_{\beta\gamma}] & = 2 f_{\alpha[\beta}{}^\delta R_{\gamma]\delta}\,.
\end{align}
In particular, this implies that the shifted generators
\begin{equation}\label{eqn:defhtau}
	\htau_\alpha = t_\alpha + R_\alpha\,,
\end{equation}
which are now completely in $\gPS$, recover the original structure coefficients through
\begin{equation}\label{eqn:commhtau}
	[ \htau_\alpha, \htau_\beta ] = - f_{\alpha\beta}{}^\gamma \htau_\gamma\,.
\end{equation}

\subsubsection{Twisted generalized torsion}\label{sec:tgT}
In the abstract view of the \PS{} construction eluded at the beginning of this section, the algebraic input is fixed by this point. Hence, it is time to get the covariant tensors it promises as output. Latter are encoded in the twisted generalized torsion. We will not derive it from scratch and instead use the results of \cite{HasslerHetPS:2024}. Still there is a bit of notation to understand before successfully applying the partially index-free results given there. At any point, it is possible to recover indices by expanding the states
\begin{equation}
	\hlangle U | = U^{\Ac} \hlangle_{\Ac}|\,, \qquad \hlangle V | = V^{\Ac} \hlangle_{\Ac}|\,, \qquad \text{and} \qquad |\partial\hrangle = |^I\hrangle\ \partial_I\,.
\end{equation}
Moreover, $|\partial V\hrangle$ denotes the partial derivative acting on $\hlangle V|$ and we adopt the convention
\begin{equation}
	\hlangle U_1 | \hlangle U_2 | A \otimes B |V_2 \hrangle |V_1 \hrangle = \hlangle U_2 | B | V_2 \hrangle \hlangle U_1 | A | V_1 \hrangle
\end{equation}
for contracting tensor products of states. All physical degrees of freedom are encoded in the frame $\cE$ which is valued in $\GPS$. However, we already know that $\GPS=\mathrm{O}(d+p,d+q)$ from \eqref{eqn:GPS}. Thus, it has to contain some auxiliary degrees of freedom which eventually can be completely expressed in terms of an $\mathrm{O}(d,d)$ valued generalized frame. Fixing all auxiliary fields lies at the heart of the gBdR identification. To help this process, we decompose
\begin{equation}\label{eqn:cEtocA}
	\boxed{\cE = \cA E \qquad \text{with} \qquad E \in \Odd}
\end{equation}
into the $\Odd$-frame $E$, and the coset element $\cA \in \GPS/\Odd$, which we aim to fix completely by torsion constraints and gauge fixing.

The expression for the twisted reduced torsion
\begin{equation}\label{eqn:THet}
	\THet_{\Ac} = \bcF_{\Ac} + \hlangle_{\Ac}| \cE |_\beta \hrangle \htau^\beta + \hlangle_{\Ac}| \htau^\beta \ZHet \cE |_\beta \hrangle
\end{equation}
is taken directly from \cite{HasslerHetPS:2024}. There, we also find that the generalized fluxes
\begin{equation}\label{eqn:cFcA}
	\bcF_{\Ac} = \genLieHet_{\scalebox{0.75}{\hlangle}_{\Ac}|\cE} \cE \cE^{-1}
\end{equation}
are computed by the heterotic generalized Lie derivative
\begin{equation}\label{eqn:hetgenLie}
	\genLieHet_{\scalebox{0.75}{\hlangle} U |} \hlangle V| = \hlangle V | \hlangle U | \partial V \hrangle + \hlangle V| \hlangle U| \ZHet |\partial U\hrangle - \hlangle V | \hlangle U | R^\alpha \otimes 1 |_\alpha\hrangle\,,
\end{equation}
with the heterotic $\ZHet$-operator
\begin{equation}\label{eqn:ZHet}
	\ZHet = 2 \Khet_{\Ac\Bc} \odot \Khet^{\Ac\Bc} = 2 K^{AB} \odot K_{AB} + R^\alpha_A \odot R_\alpha^A + \tfrac12 R^{\alpha\beta} \odot R_{\alpha\beta}
\end{equation}
where the $\odot$-product denotes the symmetric tensor product
\begin{equation}
	A \odot B = \tfrac12 ( A \otimes B + B \otimes A )\,.
\end{equation}
As already mentioned in section~\ref{sec:algebra}, we are dealing here with a special case of the heterotic \PS{} construction because $\kappa_{\alpha\beta}$ has no null-directions. Comparing with \cite{HasslerHetPS:2024}, we therefore find that all undertilded indices like $\alphan$ drop out while the tilde from the remaining indices is removed. The generators $\htau^\beta$ are the same ones we have already encountered in \eqref{eqn:defhtau}.

With the parameterization \eqref{eqn:cEtocA} of the frame $\cE$, the generalized fluxes in \eqref{eqn:cFcA} can be further simplified. To this end, we first take into account that $E$ is only generated by $K_{AB}$ and therewith satisfies
\begin{equation}\label{eqn:Ealpha}
	E |_\alpha \hrangle  = |_\alpha \hrangle\,,
\end{equation}
as can be seen from the matrix representation of its generators in \eqref{eqn:matrixKAB}. Leveraging this property of $E$, we find that the matrix elements
\begin{equation}
	\cE_{\cA\beta} = \hlangle_{\cA}| \cE |_\alpha \hrangle = \hlangle_{\cA} | \cA |_\beta \hrangle = \cA_{\cA\beta}
\end{equation}
are exclusively governed by $\cA$. Still, there are two more places where $E$ makes its appearance, namely in the flat derivatives
\begin{equation}\label{eqn:DA}
	D_A = E_A{}^I \partial_I\,, \qquad \text{with} \qquad E_A{}^I = \langle_A | E |^I \rangle\,,
\end{equation}
and the corresponding generalized fluxes
\begin{equation}\label{eqn:FABC}
	F_{ABC} = 3 D_{[A} E_B{}^I E_{C]I}\,.
\end{equation}
With the relations \eqref{eqn:Ealpha}-\eqref{eqn:FABC}, $\bcF_{\Ac}$ eventually simplifies to
\begin{equation}\label{eqn:bcF}
	\boxed{%
		\begin{aligned}
			\bcF_{\Ac} = & \hlangle_{\Ac} | \cA |^B \hrangle \left( D_B \cA \cA^{-1} + \cA \bF_B \cA^{-1} \right) - \hlangle_{\Ac} | \cA |_\beta \hrangle \cA R^{\beta} \cA^{-1} +
			\\ &\hlangle_{\cA} | D_B \cA \cA^{-1} \ZHet \cA |^B \hrangle\,,
		\end{aligned}}
\end{equation}
where $\bF_A$ denotes the generalized fluxes for the physical frame contracted with the generators $K^{BC}$,
\begin{equation}
	\bF_{A} = F_{ABC} K^{BC}\,.
\end{equation}

Although it might not be obvious at this point, \eqref{eqn:bcF} is essential to perform the first part of the gBdR identification. To already get a glimpse why, we take a look at the components of the twisted reduced torsion \eqref{eqn:THet}
\begin{equation}\label{eqn:lockingEquations}
	\THet_{\Ac\Bc\Cc} = \hlangle_{\Bc} | \THet_{\Ac} |_{\Cc} \hrangle = \cF_{\Ac\Bc\Cc} + 3 \cA_{[\Ac|\delta} (\htau^{\delta})_{|\Bc\Cc]}\,.
\end{equation}
Besides the components of the heterotic generalized fluxes, $\cF_{\Ac\Bc\Cc} = \hlangle_{\Bc}| \bcF_{\Ac} |_{\Cc}\hrangle$, the second quantity that enters here are the components of the auxiliary fields $\cA$ in the frame $\cE$. The gBdR identification arises by setting certain torsion components to zero, such that all of them are fixed in terms of the heterotic generalized fluxes \eqref{eqn:FABC} and possibly their flat derivatives \eqref{eqn:DA}. Doing this systematically requires additional information about the generalized structure group $\GS$, which we will discuss in the next section.

For completeness, note that there is also a dilatonic twisted reduced torsion. It is given by \cite{HasslerHetPS:2024}
\begin{equation}\label{eqn:THetA1}
	|\THet \hrangle = \cA |^B \hrangle F_B - D_B \cA |^B \hrangle - \htau^\beta \cA |_\beta \hrangle\,,
\end{equation}
with the dilatonic generalized flux
\begin{equation}
	F_A = 2 D_A \Phi - \partial_I E_A{}^I
\end{equation}
for the generalized dilaton $\Phi$. With the heterotic dilatonic flux
\begin{equation}\label{eqn:DHet}
	\cF_{\Ac} = \cD_{\Ac} \Phi + \partial_I \cE_{\Ac}{}^I
	\qquad \text{and} \qquad
	\cD_{\Ac} = \cE_{\Ac}{}^I \partial_I\,,
\end{equation}
\eqref{eqn:THetA1} further simplifies to
\begin{equation}\label{eqn:lockingDilatonic}
	\THet_{\Ac} = \hlangle_{\Ac} |\THet \hrangle = \cF_{\Ac} + \cA_{\Bc\gamma} (\htau^\gamma)^{\Bc}{}_{\Ac}\,.
\end{equation}
Similar to \eqref{eqn:THet}, its first term contains one derivative, while its second term is purely algebraic. Hence, it could also be used to fix at least parts of the auxiliary fields $\cA$ through torsion constraints. But for the gBdR identification this will not happen. There, it only plays a spectator role.

\subsection{Generalized structure group transformations}\label{sec:gaugeTr}
In the last subsection, we have seen that the \PS{} construction requires auxiliary fields $\cA$ besides the generalized frame $E$ on the physical space $M$. As mentioned above, these fields should be eventually fixed by imposing constraints on the torsion tensor $\THet_{\Ac}$ in \eqref{eqn:THet}. But this can only work when these constraints are invariant under the action of the generalized structure group $\GS$. Therefore, we have to figure out the transformation behavior of $E$, $\cA$ and $\THet_{\Ac}$. As instructed by \cite{HasslerHetPS:2024}, the transformations of the former two can be extracted from the quantity
\begin{equation}
	\delta_\xi \mathbb{E} = \cA^{-1} \delta_\xi \cA + \delta_\xi E E^{-1}\,.
\end{equation}
They give rise to what we call the master equation for the gauge transformations
\begin{equation}\label{eqn:trafoE}
	\boxed{%
		\delta_\xi \mathbb{E} = \left( 2 D_{[B} \xi_{A]} +  \xi^C F_{CAB} \right) K^{AB} + \xi^A \cA^{-1} D_A \cA - \xi^\alpha R_\alpha
		- D_A \xi^\beta R_\beta^A + \xi^\alpha \cA^{-1} \htau_\alpha \cA\,.
	}
\end{equation}
Looking at this relation, we see that $\Odd$-generalized diffeomorphisms are captured by the first two terms with the parameter $\xi^A$. They can be extended together with the third term to O($d+p$,$d+q$)-generalized diffeomorphisms of heterotic double field theory with the gauge group $\GS$. Based on these observations, we rewrite the transformation of $\cE$ as
\begin{equation}\label{eqn:deltaxicE}
	\delta_\xi \cE = \cA \delta_\xi \mathbb{E} E = \genLieHet_{\scalebox{0.75}{\hlangle} \xi|} \cE + \xi^{\alpha} \htau_{\alpha} \cE - D_A \xi^\beta \, \cA R^A_\beta E\,,
\end{equation}
with the heterotic generalized Lie derivative given in \eqref{eqn:hetgenLie} and the parameter $\hlangle\xi| = \hlangle_A| \xi^A + \hlangle^\alpha| \xi_\alpha$. Up to the last term, this is the transformation of the frame of heterotic double field theory under a combination of a generalized diffeomorphism and a double Lorentz transformation (assuming that $\GS$ is chosen properly).

\section{Generalized Bergshoeff-de Roo identification}\label{sec:gBdRi}
To make contact with the gBdR identification, the last thing we have to fix is the generalized structure group $\GS$. As before, it is assumed that $\kappa_{\alpha\beta}$ is non-degenerate. Hence, the simplest possible choice would be to use the double Lorentz group $\underline{\mathrm{O}(1,d-1)}\times\overline{\mathrm{O}(d-1,1)}$. However, we will see that doing so would not allow us to solve for all contributions to $\cA$ by a torsion constraint. Therefore, the structure group has to be composed more intricately, whereas the double Lorentz group only acts as a seed for a series of larger groups. Naturally, the latter will decompose into a left and a right component, like the double Lorentz group. Let us focus on the first factor because the second one will be treated in exactly the same way. We denote the series of extensions as $\underline{\widehat{\mathrm{O}}^{(p)}(1,d-1)}$, and impose the initial condition $\underline{\widehat{\mathrm{O}}^{(1)}(1,d-1)}=\underline{\mathrm{O}(1,d-1)}$. Thus, we are dealing we the generalized structure group
\begin{equation}\label{eqn:GSgBdRi}
	\GPS \supset \GS^{(p,q)} = \underline{\widehat{\mathrm{O}}^{(p)}(1,d-1)} \times \overline{\widehat{\mathrm{O}}^{(q)}(d-1,1)} = \GSL^{(p)} \times \GSR^{(q)}
\end{equation}
on an abstract level. An explicit construction is given in subsection~\ref{sec:constrGS}. To keep track of both factors, we over/underline the corresponding indices, for example
\begin{equation}
	t_{\alpha} = \begin{pmatrix} t_{\alphaL} \, & t_{\alphaR} \end{pmatrix}\,,
	\qquad \text{or} \qquad
	\THet_A = \begin{pmatrix} \THet_{\aL} \, & \THet_{\aR} \end{pmatrix}\,.
\end{equation}
The defining property of these refined indices is $\eta_{\aL\bR} = \eta_{\bR\aL} = 0$ and $\kappa_{\alphaL\betaR} = \kappa_{\betaR\alphaL} = 0$. In the following, we will show how this choice is sufficient to recover all properties of the gBdR identification.

\subsection{Parameterization of \texorpdfstring{$\cA$}{A} and partial \texorpdfstring{$\GS$}{GS}-gauge fixing}\label{sec:gaugeFixing}
In section~\ref{sec:tgT}, we argue that we eventually want to fix $\cA$ in the parameterization \eqref{eqn:cEtocA} of $\cE$ completely through torsion constraints such that the only remaining free field is the generalized frame $E$ on the physical space. Hence, a central question is:
\begin{center}
	What are the possible torsion constraints?
\end{center}
At the beginning of this section, we already imposed a left/right-split for the structure group. Also the components of the reduced twisted torsion in \eqref{eqn:lockingEquations} decompose accordingly into
\begin{equation}\label{eqn:chiralDecompThet}
	\THet_{\AcL\BcL\CcL}\,, \qquad
	\THet_{\AcL\BcL\CcR}\,, \qquad
	\THet_{\AcR\BcR\CcL}\,, \qquad \text{and} \qquad
	\THet_{\AcR\BcR\CcR}\,.
\end{equation}
For a $\THet_{\Ac\Bc\Cc}$ which transforms completely covariantly, each of these components would give rise to an independent constraint. However, as discussed in \cite{HasslerHetPS:2024} for general structure groups, $\THet_{\Ac\Bc\Cc}$ receives an inhomogeneous contribution. It is controlled by the constants entering the definition of $\hatt_{\Ah}$ in \eqref{eqn:talphadecomp}. Because these constants are completely chiral or anti-chiral for our choice of the structure group $\GS$, they only affect the totally chiral and anti-chiral components in the decomposition \eqref{eqn:chiralDecompThet}. Thus we are in general left with only the components $\THet_{\AcL\BcL\CcR}$ and $\THet_{\AcR\BcR\CcL}$ available for covariant torsion constraints. This is the first step of the gBdR identification.

This shatters the hope to fix $\cA$ purely in terms of constraints on the reduced twisted torsion because, as we just argued,  $\THet_{\AcL\BcL\CcL}$ and $\THet_{\AcR\BcR\CcR}$ are not available to fix the fully chiral, $\cA_{\AcL\BcL}$, and anti-chiral, $\cA_{\AcR\BcR}$, components of $\cA$. Instead, we have to come up with something else. Fortunately, there is another mechanism which allows to affect $\cA$, namely gauge fixing. It works by requiring that the offending fully chiral and anti-chiral contributions to $\cA$ are set to zero by appropriate $\GS$-transformations. Let us work this out more in detail. We start in full generality with the matrix exponent
\begin{equation}\label{eqn:expAp}
	\cA = \exp A'\,.
\end{equation}
Just using $A$ in the exponent would fix already the parameterization completely. But we want to have the maximal amount of freedom and therefore use odd polynomials with the expansion
\begin{equation}\label{eqn:Aprime}
	A' = A + c_1 A^3 + c_2 A^5\dots\,.
\end{equation}
Different values for the constants $c_n$ realize different parameterizations. From a physical point of view, they are all related by field redefinitions. Hence, we can in principle choose any values we like. However, choosing the correct field basis simplifies results considerably. Later we will see that $c_1 = 1/3$ results in vast cancellations and is therefore the value we will adapt. The reason why only odd terms contribute to $A'$ is that the components of $A$ have a natural $\mathbb{Z}_2$ grading in terms of their chiralities. More precisely, we have
\begin{equation}
	A = A_+ + A_- \,, \qquad \text{with} \qquad
	\left\{\begin{array}{l}
		\hlangle_{\AcL} | A_+ |^{\BcR} \hrangle = \hlangle_{\AcR} | A_+ |^{\BcL} \hrangle = 0\,, \qquad \text{and} \\
		\hlangle_{\AcL} | A_- |^{\BcL} \hrangle = \hlangle_{\AcR} | A_- |^{\BcR} \hrangle = 0\,.
	\end{array}\right.
\end{equation}
While $A_-$ will be fixed by the torsion constraints, $A_+$ is set to zero by a partial gauge fixing which breaks the generalized structure group into a subgroup. If such a gauge fixing exists is another question. We will answer it in the affirmative in section~\ref{sec:genGSTr}. Both mechanisms in combination with the right choice of the generalized structure group will fix $A$, and with it $\cA$, completely. All details are explained in the following subsections. For the moment, it is sufficient to remember $A_+$ = 0 due to gauge fixing. The same should hold for $A'$; We do not want it to contain any fully chiral/anti-chiral ($+$) components and thus are restricted to odd polynomials in $A$. This is not true for the full $\cA$ though. Here, we use the matrix exponent to obtain an element of $\GPS$ and at the same time to have an economical way to compute several quantities from the last section. In particular, we employ Hadamard's formula
\begin{equation}\label{eqn:adjointaction}
	\cA^{-1} X \cA = \sum_{m=0}^\infty \frac{(-1)^m}{m!} [ A' , X ]_m\,,
	\qquad \text{for} \qquad X \in \gPS
\end{equation}
with
\begin{equation}
	[ A, B ]_m = [ A , [ A, B ]_{m-1}]]  \,, \qquad
	[ A, B ]_0 = B
\end{equation}
to compute the adjoint action of $\cA$ in terms of nested commutators. In the same vein, we take advantage of the Baker-Campbell-Hausdorff formula
\begin{equation}\label{eqn:maurercartan}
	\cA^{-1} \delta \cA = \sum^\infty_{m=0} \frac{(-1)^m}{(m+1)!} [ A' , \delta A' ]_m\,.
\end{equation}

At this point, we are left with three major tasks:
\begin{enumerate}
	\item Fix the generalized structure group $\GS$.
	\item Fix torsion constraints which allow to find $A_-$ and solve them.
	\item Find the residual gauge transformations that are compatible with the partial gauge fixing $A_+ = 0$.
\end{enumerate}
Each of them is discussed in the following subsections.

\subsection{Construction of \texorpdfstring{$\GS$}{GS}}\label{sec:constrGS}
To proceed, we need explicit expressions for the generators $\htau_\alpha\in\gPS$ which describe $\GS$'s structure coefficients through \eqref{eqn:commhtau}. It is sufficient to discuss the details of their chiral contributions because their anti-chiral counterparts are treated in the same way. To get some motivation how to fix $\GSL$, let us look at the respective torsion constraint that follows from \eqref{eqn:lockingEquations} as
\begin{equation}
	\THet_{\AcR\BcL\CcL} = \cF_{\AcR\BcL\CcL} + \cA_{\AcR \deltaL} (\tau^{\deltaL})_{\BcL\CcL} = 0\,.
\end{equation}
For this equation to have a solution, $\tau_{\alphaL}$ has to contain all generators of $\Khet_{\AcL\BcL}$ given in \eqref{eqn:Khet}. This is very much in the spirit of the original gBdR identification. But instead of leaving these generators abstract, we reveal a new structure by using a recursive definition of $\GSL^{(p)}$. It originates from a refinement of the chiral index in
\begin{equation}
	\htau_{\alphaL} = \begin{pmatrix} \htau_{\alphaL_1} & \dots & \htau_{\alphaL_p} \end{pmatrix}
\end{equation}
and is based on the identifications
\begin{equation}\label{eqn:splittingfun}
	\begin{aligned}
		\htau_{\alphaL_1}     & = \begin{pmatrix} \htau^{\aL\bL} \end{pmatrix}\,,                                                                                        \\
		\htau_{\alphaL_2}     & = \begin{pmatrix} \htau^{\aL}_{\betaL_1} \quad & \htau_{\alphaL_1\betaL_1} \end{pmatrix}\,, \qquad \text{and from there on}              \\
		\htau_{\alphaL_{i+1}} & = \begin{pmatrix} \htau^{\aL}_{\betaL_i} \quad & \htau_{\alphaL_1\betaL_i} \quad & \dots \quad & \htau_{\alphaL_i\betaL_i} \end{pmatrix}
	\end{aligned}
\end{equation}
until we reach $\htau_{\alphaL_p}$. To make sense of these relations, we have to express the new generators $\htau^{\aL\bL}$, $\htau^{\aL}_{\betaL_i}$ and $\htau_{\alphaL_i\betaL_j}$ that appear or the right-hand side in terms of generators we already know. We start with the Lorentz generators
\begin{equation}\label{eqn:tau00}
	\htau^{\aL\bL} = g_- \, K^{\aL\bL},
\end{equation}
where $g_-$ is an overall normalization constant which is eventually expressed in terms of the $a$ used in the gBdR identification of \cite{Baron:2020xel}. This is a natural choice because it implies
\begin{equation}
	\GSL^{(1)} = \underline{\mathrm{O}(1,d-1)} \,.
\end{equation}
The remaining two generators are fixed by the relation
\begin{equation}
	\Khet^{\AcL\BcL} = \frac{1}{g_-}\begin{pmatrix}
		\phantom{-}\htau^{\aL\bL}\quad & \htau^{\aL}_{\betaL}  \\
		-\htau^{\bL}_{\alphaL}\quad    & \htau_{\alphaL\betaL}
	\end{pmatrix}\,.
\end{equation}
In addition to \eqref{eqn:tau00}, this gives eventually rise to
\begin{equation}\label{eqn:tauij}
	\htau^{\aL}_{\betaL_i} = \frac{g_-}{2} \, R^{\aL}_{\betaL_i}\,,
	\qquad \text{and} \qquad
	\htau_{\alphaL_i\betaL_j} = - \frac{g_-}{2} \, R_{\alphaL_i\betaL_j}\,.
\end{equation}
At this point, we see why it is more convenient to have the generators $\htau_{\alphaL_i\betaL_j}$ with two indices instead of one. Counting them gives rise to the dimension
\begin{equation}
	\dim \GSL^{(p+1)} = \tfrac12 \left( d + \dim \GSL^{(p)} \right)\left( d + \dim \GSL^{(p)} - 1 \right)\,,
\end{equation}
with the initial condition
\begin{equation}
	\dim \GSL^{(0)} = 0\,.
\end{equation}
Although this dimension grows exponentially with $p$, it is finite for any finite $p$. Hence, in contrast to the standard approach to the gBdR identification \cite{Baron:2018lve,Baron:2020xel}, we do not need to deal a priori with infinite-dimensional Lie algebras. Our organization of generators suggests the hierarchy of subgroups
\begin{equation}\label{eqn:GSHier}
	\GSL^{(1)} \subset \dots \subset \GSL^{(p)} \,.
\end{equation}

In order to prove this conjecture, we first have to fix $\kappa_{\alpha\beta}$, because it appears in the relevant commutators \eqref{eqn:commR1R1}, \eqref{eqn:commR1R2} and \eqref{eqn:commR2R2}. Without loss of generality, we impose
\begin{equation}\label{eqn:kappadiag}
	\kappa_{\alphaL_i \betaL_j} = 0 \qquad \text{for} \qquad i \ne j\,.
\end{equation}
This might be seen as the first step towards choosing a basis for $\htau_{\alpha}$ where $\kappa_{\alpha\beta}$ is diagonal. To further analyze the properties of $\GSL^{(l)}$, we need the commutators of its generators. To write them in a simple form, we extend the index ${}_{\alphaL_i}$ by including ${}_{\alphaL_0} = {}^{\aL}$. Thereby, \eqref{eqn:tau00} and \eqref{eqn:tauij} are unified into $\tau_{\alpha_I\beta_J}$, where capital indices like $I$ and $J$ start with zero. With this convention, the commutators we look for take on the simple form
\begin{equation}\label{eqn:commtaudoubleind}
	[ \htau_{\alphaL_I\betaL_J}, \htau_{\gammaL_K\deltaL_L} ] = 2 g_- \, \heta_{[\alphaL_I|[\gammaL_K} \htau_{\deltaL_L]|\betaL_J]}\,,
\end{equation}
which follows immediately from \eqref{eqn:commKhetKhet}. The property \eqref{eqn:kappadiag} carries over to $\heta_{\alphaL_i\betaL_j}$ such that we find
\begin{equation}
	\heta_{\alphaL_i\betaL_j} = 0 \qquad\text{for}\qquad i \ne j\,.
\end{equation}
Because $\GSL^{(l)}$ is generated by all $\htau_{\alphaL_I\betaL_J}$ with $0 \le I \le J \le l-1$, it closes under the commutators \eqref{eqn:commtaudoubleind}. Thereby, we confirm the hierarchy of subgroups anticipated in \eqref{eqn:GSHier}. A comparison with \eqref{eqn:commKABKCD} furthermore reveals that $\heta^{\AcL\BcL}$ is the invariant pairing of $\GSL^{(l)}$.

For later, we also need an explicit expression for the structure coefficients $f_{\alpha\beta}{}^\gamma$. They can be read off directly from \eqref{eqn:commtaudoubleind}, resulting in
\begin{equation}\label{eqn:f6explicit}
	f_{\alphaL_I \betaL_J \gammaL_K \deltaL_L}{}^{\epsilonL_M \rhoL_N} = - 2 g_- \heta_{[\alphaL_I|[\gammaL_K} \delta_{\deltaL_L]}^{[\epsilonL_M} \delta_{|\betaL_J]}^{\rhoL_N]} \,.
\end{equation}
All further information is completely encoded in $\kappa_{\alphaL_i\betaL_i}$ which can be computed recursively from the invariant pairing on $\GPS$ induced by $\ZHet$. More precisely, consider the pairing
\begin{equation}\label{eqn:pairingExplicit}
	\begin{aligned}
		\llangle K_{AB}, K_{CD} \rrangle                    & = \eta_{[A|[C} \eta_{D]|B]} \,,                         \\
		\llangle R_\alpha^A , R_\beta^B \rrangle            & = 2 \eta^{AB} \kappa_{\alpha\beta}\,, \qquad \text{and} \\
		\llangle R_{\alpha\beta}, R_{\gamma\delta} \rrangle & = 4 \kappa_{[\alpha|[\gamma} \kappa_{\delta]|\beta]}\,.
	\end{aligned}
\end{equation}
It is normalized such that
\begin{align}
	\llangle \ZHet , X \rrangle = 2 X \qquad                                                        & \text{for all} \qquad X \in \gPS
	\intertext{holds and of course it also satisfies}
	\llangle [ \htau_\alpha, X ], Y \rrangle +   \llangle X, [ \htau_\alpha, Y ] \rrangle = 0\qquad & \text{for all} \qquad X, Y \in \gPS\,,
\end{align}
the index free version of \eqref{eqn:kappainvaraint}. We use it to compute
\begin{equation}\label{eqn:pairing}
	\frac{1}{g_-^{2}} \llangle \htau_{\alphaL_I\betaL_J}, \htau_{\gammaL_K \deltaL_L} \rrangle = \heta_{[\alphaL_I|[\gammaL_K} \heta_{\deltaL_L]|\betaL_J]} = \kappa_{\alphaL_I \betaL_J \gammaL_K \deltaL_L}\,.
\end{equation}
Looking at \eqref{eqn:pairingExplicit}, one sees that this definition indeed is compatible with our initial requirement \eqref{eqn:kappadiag}. As it is used to raise and lower Greek indices, we do not want $\kappa_{\alpha\beta}$ to depend on any parameters. Hence, the factor $1/g_-^2$ is added on the left-hand side. At the leading order, we obtain
\begin{equation}
	\kappa_{\alphaL_1\betaL_1} = \kappa_{\aL_1\aL_2\bL_1\bL_2} = \eta_{[\aL_1| [\bL_1} \eta_{\bL_2]|\aL_2]}
\end{equation}
after expanding the indices. From there on, $\kappa_{\alphaL_i\betaL_i}$ is computed recursively from $\kappa_{\alphaL_{i-1}\betaL_{i-1}}$, \dots, $\kappa_{\alphaL_1\betaL_1}$. As an example, take
\begin{equation}
	\kappa{}_{\alphaL_2\betaL_2} = \begin{pmatrix}
		\kappa^{\aL_1}_{\aL_2\aL_3}{}^{\bL_1}_{\bL_2\bL_3} & 0                                                 \\
		0                                                  & \kappa_{\aL_1\aL_2\aL_3\aL_4\bL_1\bL_2\bL_3\bL_4}
	\end{pmatrix},
\end{equation}
with
\begin{align}
	\kappa^{\aL_1}_{\aL_2\aL_3}{}^{\bL_1}_{\bL_2\bL_3} & = \tfrac12 \, \eta^{\aL_1\bL_1} \eta_{[\aL_2|[\bL_2} \eta_{\bL_3]|\aL_3]} \,, \qquad \text{and}                                                                                                     \\
	\kappa_{\aL_1\aL_2\aL_3\aL_4\bL_1\bL_2\bL_3\bL_4}  & = \tfrac12 \bigl(\eta_{[\aL_1 | [\bL_1}\eta_{\bL_2]|\aL_2]}\eta_{[\bL_3 |[\aL_3}\eta_{\aL_4]|\bL_4]} - \eta_{[\aL_1 | [\bL_3}\eta_{\bL_4]|\aL_2]}\eta_{[\bL_1 |[\aL_3}\eta_{\aL_4]|\bL_2]}\bigr)\,.
\end{align}

In the following, we will see that a complete identification of $A$ for the gauge group $\GSL^{(p)}$ requires the extension to $\GSL^{(p+1)}$. This process repeats again and again. Initially, it looks like it can be truncated if the maximal number of derivatives is limited. However, we will point out a problem for finite dimensional structure groups in the next subsection. The discussion for the anti-chiral sector is nearly identical; The only difference is that we swap $g_-$ with a second normalization constant $g_+$. Later on, we will see that in this way we recover the bi-parametric deformation that is captured by the gBdR identification. In conclusion, we are indeed dealing with the decomposition of the generalized structure group anticipated in \eqref{eqn:GSgBdRi}.

\subsection{Collapsing towers}\label{sec:collapsingTowers}
We have now constructed a simple basis for $\GSL^{(n)}$ in terms of the generators $\htau_{\alphaL}$, which will play the central role in the \PS{} construction. However, the torsion constraints and the gauge-fixing in the next two subsections are rather written in terms of the generators $t_{\alphaL}$. Both are related by the change of basis
\begin{equation}\label{eqn:transformS}
	t_{\alphaL} = S_{\alphaL}{}^{\betaL} \htau_{\betaL}\,,
\end{equation}
with
\begin{equation}
	S_{\alphaL}{}^{\betaL} =  \delta_{\alphaL}{}^{\betaL} - \phi_{\alphaL}{}^{\betaL}\,,
\end{equation}
and
\begin{equation}\label{eqn:defphi}
	\phi_{\alphaL}{}^{\betaL} = \tfrac{1}{g_-} f_{\alphaL}{}^{\betaL'\betaL''}\,.
\end{equation}
To make sense of the last equation, note that we split the double index $\betaL$ into its components denoted by $\betaL'$ and $\betaL''$ by using \eqref{eqn:splittingfun}. We will encounter this technique of splitting indices very often in the following. Although the relation \eqref{eqn:transformS} looks quite innocent at the beginning, it causes a lot of work in the identification. The main reason is that we are not dealing with the map $S$ itself in the identification but with its inverse $S^{-1}$, which is more complicated and is defined by
\begin{equation}
	(S^{-1})_{\alphaL}{}^{\gammaL} S_{\gammaL}{}^{\betaL} = S_{\alphaL}{}^{\gammaL} (S^{-1})_{\gammaL}{}^{\betaL} = \delta_{\alphaL}^{\betaL}\,.
\end{equation}
It is given by the sum
\begin{equation}\label{eqn:Sinv}
	(S^{-1})_{\alphaL}{}^{\betaL} = \sum_{n=0}^{p-1} (\phi^n)_{\alphaL}{}^{\betaL}\,,
	\qquad \text{where} \qquad
	(\phi^0)_{\alphaL}{}^{\betaL} = \delta_{\alphaL}^{\betaL}
\end{equation}
of higher and higher powers in $\phi$. From the definition \eqref{eqn:defphi}, we see that when $\phi$ is applied to elements in the algebra  $\underline{\gs}^{(n)}$ it results in an element of $\underline{\gs}^{(n+1)}$. Therefore, we conclude that
\begin{equation}
	\phi^m: \underline{\gs}^{(n)} \to \underline{\gs}^{(n+m)}
\end{equation}
and thereby understand that even if $S^{-1}$ is applied to very simple generators, like the left-handed double Lorentz generators of $\underline{\GS}^{(1)}$, it gives a whole tower of generators filling $\underline{\gS}^{(p)}$. Although most of the generators in the tower are redundant, we still have to deal with them. This is the main reason why the gBdR identification is so complicated.

At a first step towards a better understanding of these towers' fate, consider the situation where two of them are contracted by the pairing in \eqref{eqn:pairing}. This is a situation that will occur frequently in the identification, motivating us to introduce
\begin{equation}\label{eqn:defkappatilde}
	\tkappa_{\alpha\beta} :=  (S^{-1})_{\alphaL}{}^{\gammaL}\, \kappa_{\gammaL\deltaL}\, (S^{-1})_{\betaL}{}^{\deltaL}\,.
\end{equation}
All we need to know in order to compute this quantity is
\begin{equation}\label{eqn:phicontract}
	\phi_{\alphaL}{}^{\gammaL} \kappa_{\gammaL\deltaL}^{(m)} \phi_{\betaL}{}^{\deltaL} = X_m \, \kappa^{(m-1)}_{\alphaL\betaL}\,,
\end{equation}
with the Dynkin index
\begin{equation}\label{eqn:dynkinindex}
	X_m = d + \dim \GSL^{(m-1)} - 2
\end{equation}
and the restricted pairing
\begin{equation}
	\kappa^{(m)}_{\alpha_i\beta_j} = \begin{cases}
		\kappa_{\alpha_i\beta_j} & \text{for }i,j\le m\,, \\
		0                        & \text{otherwise}\,.
	\end{cases}
\end{equation}
After applying \eqref{eqn:phicontract} recursively, we are able to evaluate the sum
\begin{equation}
	\upkappa_{\alphaL\betaL} := \sum\limits_{n=0}^{p-1} (\phi^n)_{\alphaL}{}^{\gammaL} \kappa^{(p)}_{\gammaL\deltaL} (\phi^n)_{\betaL}{}^{\deltaL}
\end{equation}
as
\begin{equation}
	\upkappa_{\alphaL_i\betaL_j} = \begin{cases}
		\chiL_i \kappa_{\alphaL_i\betaL_i} & \text{for } i = j\,, \\
		0                                  & \text{otherwise}
	\end{cases}
\end{equation}
with constants $\chi_i$ given by
\begin{equation}
	\chiL_m = 1 + \sum_{n=1}^{p-m} \prod_{l=1}^n X(p-l)\,.
\end{equation}
After this preparation, we can finally simplify \eqref{eqn:defkappatilde} to
\begin{equation}\label{eqn:kappaCollapsed}
	\tkappa_{\alphaL\betaL} = \upkappa_{\alphaL\betaL} + 2 (\tSinv)_{(\alphaL}{}^{\gammaL} \upkappa_{\betaL)\gammaL}\,,
\end{equation}
where we have introduced for convenience
\begin{equation}\label{eqn:tSinv}
	(\tSinv)_{\alphaL}{}^{\betaL} = (S^{-1})_{\alphaL}{}^{\betaL} - \delta_{\alphaL}^{\betaL}\,.
\end{equation}
It also follows from \eqref{eqn:Sinv} that
\begin{equation}\label{eqn:tSinvtoSinv}
	(\tSinv)_{\alphaL}{}^{\betaL}=(S^{-1})_{\alphaL}{}^{\gammaL}\phi_{\gammaL}{}^{\betaL}\,.
\end{equation}
We have now successfully collapsed the towers contributing to the original definition \eqref{eqn:defkappatilde}. Only the constants $\chiL_i$ remain as a reminder of the redundancy generated by $S^{-1}$.

In the same way we discovered the relation between $\tkappa_{\alpha\beta}$ and $\kappa_{\alpha\beta}$, we need to deal with the structure coefficients. As starting point here one may take the commutator
\begin{equation}
	- [ t_{\alphaL}, t_{\betaL} ] = S_{\alphaL}{}^{\deltaL} S_{\betaL}{}^{\epsilonL} f_{\deltaL\epsilonL}{}^{\gammaL} \tau_{\gammaL} = f_{\alphaL\betaL}{}^{\gammaL} \tau_{\gammaL} + 2 [ R_{[\alphaL}, \tau_{\betaL]} ] + f_{\alphaL\betaL}{}^{\gammaL} R_{\gammaL}\,.
\end{equation}
Using
\begin{equation}
	[ R_{\alphaL}, \tau_{\betaL} ] = - \phi_{\alphaL}{}^{\deltaL} f_{\deltaL\betaL}{}^{\gammaL} \tau_{\gammaL}\,, \qquad\text{and}\qquad
	R_{\alphaL} = \phi_{\alphaL}{}^{\betaL} \tau_{\betaL}\,,
\end{equation}
we then find
\begin{equation}\label{eqn:SSf}
	S_{\alphaL}{}^{\deltaL} S_{\betaL}{}^{\epsilonL} f_{\deltaL\epsilonL}{}^{\gammaL} = 2 S_{[\alphaL|}{}^{\deltaL} f_{\deltaL|\betaL]}{}^{\gammaL} - f_{\alphaL\betaL}{}^{\deltaL} S_{\deltaL}{}^{\gammaL}\,.
\end{equation}
Multiplying it with $(S^{-1})_{\alphaL}{}^{\gammaL}(S^{-1})_{\betaL}{}^{\deltaL}$, along with using
\begin{equation}
	S_{\alphaL}{}^{\gammaL}\tkappa_{\gammaL\betaL} = (S^{-1})_{\betaL\alphaL}
\end{equation}
from \eqref{eqn:defkappatilde}, we obtain the counterpart of $\tkappa_{\alphaL\betaL}$,
\begin{equation}\label{eqn:SinvSinvf}
	\tf_{\alphaL\betaL\gammaL} :=  (S^{-1})_{\alphaL}{}^{\deltaL}(S^{-1})_{\betaL}{}^{\epsilonL}(S^{-1})_{\gammaL}{}^{\rhoL}f_{\deltaL\epsilonL\rhoL} = 2(\tSinv)_{[\alphaL|}{}^{\deltaL}f_{\deltaL|\betaL]}{}^{\epsilonL}\tkappa_{\epsilonL\gammaL} +f_{\alphaL\betaL}{}^{\deltaL}\tkappa_{\deltaL\gammaL}\,.
\end{equation}

At the end of the last subsection, it was concluded that in order to fix all components of $\cA$ one has to send the $p$ of $\GSL^{(p)}$ eventually to $\infty$. This will make the initial point of our analysis in \eqref{eqn:phicontract} more subtle because we have to regularize the expression for the divergent Dynkin index in \eqref{eqn:dynkinindex}. There are different ways to do so. Perhaps the most obvious way would be to just keep \eqref{eqn:phicontract} and consider instead of $\chi_i$ the ratio $\chi_i/\chi_1$ that remains finite in the limit $p\rightarrow\infty$. However, there is an alternative regularization based on the observation that the difference between $\kappa^{(m)}_{\gammaL\deltaL}$ and $\kappa^{(m-1)}_{\alphaL\betaL}$ on both sides of \eqref{eqn:phicontract} only appears because we are working with a finite structure group. But for $p\rightarrow\infty$, it is more natural to substitute \eqref{eqn:phicontract} by
\begin{equation}\label{eqn:phicontractBetter}
	\phi_{\alphaL}{}^{\gammaL} \kappa_{\gammaL\deltaL} \phi_{\betaL}{}^{\deltaL} = X_\infty \, \kappa_{\alphaL\betaL}\,.
\end{equation}
Remarkably, this choice simplifies matters considerably because using \eqref{eqn:phicontractBetter} instead of \eqref{eqn:phicontract}, one finds that all $\chiL_i$ are equal, implying
\begin{equation}\label{eqn:correctReg}
	\boxed{\chiL_i = \chiL\,.}
\end{equation}
Of course this remaining $\chiL$ is infinite, but as in the original gBdR identification, it can be combined with the constant $g_-$ to form the finite parameter $a$. In the beginning, we have been hesitant to commit to an infinite $\GS$ but now we showed that instead of making matters more complicated, it actually simplifies the collapsing of towers considerably. On one hand, it might be worrisome to have to choose between different regularization procedures. On the other hand, this situation is well known from quantum field theory. There, the regularization has to be performed in a way that is compatible with the symmetries of the theory. Here, we will make the same observation: Only with the regularization prescription \eqref{eqn:correctReg} the residual gauge transformation that survives after the gauge fixing in section~\ref{sec:genGSTr} will close.

With the simplifications introduced by \eqref{eqn:correctReg}, we can for example compute $\tilde{f}_{\alphaL_1\betaL_1\gammaL_1}$ which will be needed later. As the first step, we note that \eqref{eqn:kappaCollapsed} simplifies to
\begin{equation}\label{eqn:tkappatokappa}
	\tkappa_{\alphaL\betaL} = \chiL \kappa_{\alphaL\betaL} + 2\chiL (\tSinv)_{(\alphaL\betaL)}\,,
\end{equation}
allowing us to rewrite \eqref{eqn:SinvSinvf} as
\begin{equation}
	\tilde{f}_{\alphaL\betaL\gammaL} = 2 \chiL (\tSinv)_{[\alphaL|}{}^{\deltaL}f_{\deltaL|\betaL]\gammaL}+4 \chiL (\tSinv)_{[\alphaL|}{}^{\deltaL}f_{\deltaL|\betaL]}{}^{\epsilonL}(\tSinv)_{(\epsilonL\gammaL)} + \chiL f_{\alphaL\betaL\gammaL} + 2\chiL f_{\alphaL\betaL}{}^{\deltaL}(\tSinv)_{(\deltaL\gammaL)}\,.
\end{equation}
Furthermore, from \eqref{eqn:SinvSinvf} and \eqref{eqn:tSinv} one finds
\begin{equation}\label{eqn:tstsf}
	(\tSinv)_{\alphaL}{}^{\deltaL}(\tSinv)_{\betaL}{}^{\epsilonL}f_{\deltaL\epsilonL\gammaL} = 2(\tSinv)_{[\alphaL|}{}^{\delta}f_{\deltaL|\betaL}{}^{\epsilonL}(\tSinv)_{\epsilonL\gammaL} + f_{\alphaL\betaL}{}^{\deltaL}(\tSinv)_{\deltaL\gammaL}
\end{equation}
and finally obtains
\begin{equation}
	\begin{aligned}\label{eqn:tf}
		\tilde{f}_{\alphaL\betaL\gammaL} = & 2\chiL (\tSinv)_{[\alphaL|}{}^{\deltaL}f_{\deltaL|\betaL]\gammaL} + \chiL f_{\alphaL\betaL}{}^{\deltaL}(\tSinv)_{\gammaL\deltaL} + 6 \chiL (\tSinv)_{[\alphaL|}{}^{\deltaL}f_{\deltaL|\betaL|}{}^{\epsilonL}(\tSinv)_{\epsilonL|\gammaL]} + \\ &3\chiL f_{[\alphaL\betaL|}{}^{\deltaL}(\tSinv)_{\deltaL|\gammaL]} + \chiL f_{\alphaL\betaL\gammaL}\,.
	\end{aligned}
\end{equation}
Restricting the indices to $\alpha_1$, $\beta_1$ and $\gamma_1$, one deduces that
\begin{equation}
	\tilde{f}_{\alphaL_1\betaL_1\gammaL_1} = \chiL f_{\alphaL_1\betaL_1\gammaL_1}
\end{equation}
because $(\tSinv)_{\alphaL_1}{}^{\betaL_1} = 0$ as can be seen from \eqref{eqn:tSinv}. Another quantity that we will need later is
\begin{equation}\label{eqn:tftktf}
	(S^{-1})_{\alphaL_1}{}^{\muL}(S^{-1})_{\betaL_1}{}^{\nuL}f_{\muL\nuL\gammaL}(S^{-1})_{\deltaL_1}{}^{\rhoL}(S^{-1})_{\epsilonL_1}{}^{\sigmaL}f_{\rhoL\sigmaL}{}^{\gammaL} = \tf_{\alphaL_1\betaL_1\gammaL}\tf_{\deltaL_1\epsilonL_1\rhoL}\tkappa^{\gammaL\rhoL}\, ,
\end{equation}
where \eqref{eqn:tf} and \eqref{eqn:defkappatilde} were used to obtain the right-hand side. With the aid of \eqref{eqn:SinvSinvf}, \eqref{eqn:tkappatokappa} and \eqref{eqn:tstsf}, it is straightforward to obtain
\begin{equation}\label{eqn:chiff}
	\tf_{\alphaL_1\betaL_1\gammaL}\tf_{\deltaL_1\epsilonL_1\rhoL}\tkappa^{\gammaL\rhoL} = \chiL f_{\alphaL_1\betaL_1\gammaL_1}f_{\deltaL_1\epsilonL_1}{}^{\gammaL_1}\,.
\end{equation}
Of course all of these relations hold for the conjugate chiralities as well, by swapping $\chiL$ with $\chiR$.

\subsection{Identification}\label{sec:identification}
Next, we would like to fix $A$, and with it $\cA$, by coming back to the idea of setting selected reduced twisted torsion components in \eqref{eqn:lockingEquations} to zero. Due to \eqref{eqn:cEtocA}, it is possible to replace $\cE$ with $\cA$ in $\THet_{\cA}$. Moreover, we only need to consider the mixed chirality components $\cA_{\aL\betaR}$, $\cA_{\alphaL \betaR}$, $\cA_{\aR \betaL}$, and $\cA_{\alphaR \betaL}$ because chiral/anti-chiral contributions are set to zero by partial gauge fixing as explained in a previous subsection. Consequentially, we are left with the four constraints
\begin{align}\label{eqn:lock1R}
	0 & = \THet_{\aL\BcR\CcR}          &  & \longleftrightarrow & \cA_{\aL\deltaR} (\htau^{\deltaR})_{\BcR\CcR}          & = - \cF_{\aL\BcR\CcR}\,,                            \\\label{eqn:lock1L}
	0 & = \THet_{\aR\BcL\CcL}          &  & \longleftrightarrow & \cA_{\aR\deltaL} (\htau^{\deltaL})_{\BcL\CcL}          & = - \cF_{\aR\BcL\CcL}\,,                            \\\label{eqn:lock2R}
	0 & = \THet^{\alphaL}{}_{\BcR\CcR} &  & \longleftrightarrow & \cA^{\alphaL}{}_{\deltaR} (\htau^{\deltaR})_{\BcR\CcR} & = - \cF^{\alphaL}{}_{\BcR\CcR}\,, \qquad \text{and} \\\label{eqn:lock2L}
	0 & = \THet^{\alphaR}{}_{\BcL\CcL} &  & \longleftrightarrow & \cA^{\alphaR}{}_{\deltaL} (\htau^{\deltaL})_{\BcL\CcL} & = - \cF^{\alphaR}{}_{\BcL\CcL}\,,
\end{align}
from \eqref{eqn:lockingEquations} with $(\htau^\alpha)_{\Bc\Cc} = \hlangle_{\Bc} | \htau^\alpha |_{\Cc} \hrangle$.  Here we observe the pattern that was already anticipated at the end of section~\ref{sec:constrGS}: The right-hand sides of \eqref{eqn:lock1R} to \eqref{eqn:lock2L} restricted to $\GSL^{(p)}$ will require the extension to $\GSL^{(p+1)}$ to satisfy the torsion constraints. The new components of $\cA$ which are generated in this step can be written exclusively in terms of the generalized frame on the physical space $E$ and at least $p+1$ derivatives.

We extract the relevant components of $A$ by using the pairing introduced in \eqref{eqn:pairing}. This is done iteratively in the number of derivatives, which we will keep track of by decorating the respective quantities like $\cF_{\Ac}$ by $\cF_{\Ac}^{(l)}$ when they contain $l$ derivatives. But, due to the second term of \eqref{eqn:bcF}, one encounters
\begin{equation}
	\cA^{(l)\beta}_{\Ac} \htau_\beta \cong - \bcF^{(l)}_{\Ac} = \cA^{(l)\beta}_{\Ac} R_\beta - \bcF^{(l)}_{\Ac}[A^{(<l)}]\,.
\end{equation}
Here, we introduce the equivalence relation $\cong$ which ignores all contributions that are not contained in $\cA^{(l)}_{\Ac}{}^\beta$, like generators with mixed chirality. In the following, we will use it frequently to avoid the need to write many irrelevant terms. Moreover, we indicate with $\bcF^{(l)}_{\Ac}[A^{(<l)}]$ all contributions to $\bcF^{(l)}$ which contain $A$ at most up to order $A^{(l-1)}$. To achieve a separation of orders on the left- and right-hand side, which is needed to eventually solve for $A^{(l)}$ iteratively, we equivalently write
\begin{equation}\label{eqn:identStep1}
	\cA^{(l)\beta}_{\Ac} t_\beta \cong - \bcF^{(l)}_{\Ac} [ A^{(<l)}]\,.
\end{equation}
At this point, we re-encounter the generator $t_\alpha$ which we have introduced in section~\ref{sec:GS}. Finally, we transition from $\cA$ to $A$ by using the expansion \eqref{eqn:Aprime}, leading to
\begin{equation}
	A^{(l)\beta}_{\Ac} t_{\beta} \cong - \bcF^{(l)}_{\Ac} [ A^{(<l)} ] - \bcG^{(l)\beta}_{\cA} [ A^{(<l)} ] t_{\beta}
\end{equation}
with
\begin{equation}
	\bcG^{(l)\beta}_{\cA} [ A^{(<l)} ] = \cA^{(l)\beta}_{\Ac} -  A^{(l)\beta}_{\Ac}\,.
\end{equation}
A direct computation shows us
\begin{align}
	\bcG^{(1)}_{\cA\beta} & = 0 \,,                                                                              \\
	\bcG^{(2)}_{\cA\beta} & = \tfrac12 \hlangle_{\Ac} | (A^{(1)})^2 |_\beta \hrangle \cong 0\,,
	\intertext{and the first non-trivial contribution}
	\bcG^{(3)}_{\cA\beta} & \cong \left( c_1 + \tfrac16 \right) \hlangle_{\Ac} | (A^{(1)})^3 |_\beta \hrangle\,.
\end{align}
We rather work with the generators $\htau_\alpha$ because they form the natural basis for $\GS^{(p)}$ introduced in the last subsection. Therefore, we define
\begin{align}\label{eqn:AtildeToA}
	A^{(l)}_{\Ac}{}^\beta t_\beta & = \tA^{(l)}_{\Ac}{}^\beta \htau_{\beta}\,,                &
	\bcG_{\Ac}^{(l)\beta} t_\beta & = \tbcG_{\Ac}^{(l)\beta} \htau_{\beta}\,,
	\intertext{with}
	A^{(l)}_{\Ac}{}^\beta         & = \tA^{(l)}_{\Ac}{}^\gamma (S^{-1})_{\gamma}{}^{\beta}\,, &
	\bcG_{\Ac}^{(l)\beta}         & = \tbcG_{\Ac}^{(l)\gamma} (S^{-1})_{\gamma}{}^{\beta}\,,
\end{align}
where $S^{-1}$ is given in \eqref{eqn:Sinv}. At this point, the idea of collapsing towers explain in section~\ref{sec:collapsingTowers} becomes relevant. As explained there, compared to the untilded version, $\tA_{\Ac\beta}$ has the advantage that it has a finite number of contributions. Its components arise from the identification
\begin{equation}\label{eqn:identA}
	\boxed{%
		\begin{aligned}
			\tA^{(l)}_{\AcR\betaL} & = - \frac{1}{g_-^2} \llangle \bcF^{(l)}_{\AcR}[A^{(<l)}],  \htau_{\betaL} \rrangle - \tbcG^{(l)}_{\AcR\betaL}[A^{(<l)}]\,, \\
			\tA^{(l)}_{\AcL\betaR} & = - \frac{1}{g_+^2} \llangle \bcF^{(l)}_{\AcL}[A^{(<l)}] , \htau_{\betaR} \rrangle - \tbcG^{(l)}_{\AcL\betaR}[A^{(<l)}]\,.
		\end{aligned}}
\end{equation}

To better understand how to use \eqref{eqn:identA}, let us perform the identification at the leading order. Starting point is $\bcF^{(0)}_{\cA}$ which only has a contribution from
\begin{equation}\label{eqn:F0}
	\bcF^{(0)\alpha} =- R^\alpha  \cong 0\,.
\end{equation}
$R^\alpha$ in the middle of this relation is purely chiral or anti-chiral and one thus has $A^{(0)}=0$; There are no contributions from $A$ at this order. Fortunately, they are not needed because, as \eqref{eqn:F0} shows, the relevant components of $\bcF^{(0)}$ vanish on their own. Things become more interesting at the next order. Here one encounters
\begin{equation}
	\bcF_A^{(1)} = \bF_A \,, \qquad \text{and} \qquad
	\bcF^{(1)\alpha} = 0\,.
\end{equation}
With this information, we get
\begin{align}\label{eqn:identA1}
	K: &  & \tA^{(1)}_{\aR\betaL_1} & = \tA^{(1)\bL_1\bL_2}_{\aR} = - \tfrac{1}{g_-} F_{\aR}{}^{\bL_1\bL_2}\,,
	\intertext{and}
	K: &  & \tA^{(1)}_{\aL\betaR_1} & = \tA^{(1)\bR_1\bR_2}_{\aL} = - \tfrac{1}{g_+} F_{\aL}{}^{\bR_1\bR_2}\,.
\end{align}
Here $K:$ is a reminder that only generators $\tau^{AB} \sim K^{AB}$ contribute in this identification. For higher orders, we will also encounter $\tau^A_\beta$ and $\tau_{\alpha\beta}$. As they correspond to $R$-generators, we will denote their contributions by $R_1$, and $R_2$ respectively. In the computation of higher orders, the index-free version
\begin{equation}\label{eqn:defA1}
	A^{(1)} = A^{(1)}_{\aR \betaL} R^{\aR\betaL} + A^{(1)}_{\aL \betaR} R^{\aL\betaR}
\end{equation}
will prove very handy. This can be seen already at the next order, where we face the relevant parts
\begin{align}\label{eqn:F2A}
	\bcF_A^{(2)} & \cong [ A^{(1)}, \bF_A ] + \hlangle_A | D_B A^{(1)} \ZHet |^B \hrangle  =  \left( F_{ACB} A^{(1)C}{}_\gamma - D_B A^{(1)}_{A \gamma} \right) R^{B \gamma}
	\,, \qquad \text{and}                                                                                                                                                         \\	\bcF^{(2) \alpha} & \cong \hlangle^\alpha | A^{(1)} |^B \hrangle \bF_B - \tfrac12 [ A^{(1)}, [ A^{(1)} , R^\alpha] ] + \hlangle^\alpha| D_B A^{(1)} \ZHet |^B \hrangle \nonumber \\
	             & = \left( 2 D_B A^{(1)}_C{}^\alpha - A^{(1)}_D{}^\alpha F^D{}_{BC} - f^{\alpha\beta\gamma} A^{(1)}_{B\beta} A^{(1)}_{C\gamma} \right) K^{BC}\label{eqn:F2alpha}
\end{align}
of the heterotic fluxes. Note that in expanding \eqref{eqn:bcF}, one initially gets $A'$ after applying \eqref{eqn:adjointaction} and \eqref{eqn:maurercartan}. However, at the leading order we find
\begin{equation}
	A'^{(1)} = A^{(1)}\,\qquad \text{and} \qquad
	A'^{(2)} = A^{(2)}
\end{equation}
and therefore can just drop the prime. The $R_1$-component from \eqref{eqn:F2A} gives rise to the identification
\begin{align}\label{eqn:A2a}
	R_1: &  & \tA^{(2)}_{\aR}{}^{\bL}{}_{\betaL} & = \tfrac{1}{g_-} \left( F_{\aR}{}^{\bL \cR} A^{(1)}_{\cR\betaL} + D^{\bL} A^{(1)}_{\aR\betaL} \right)\,.
	\intertext{In the same vein, the $K$-contribution from \eqref{eqn:F2alpha} results in}
	\label{eqn:A2alphaRbetaL}
	K:   &  & \tA_{\alphaR}^{(2)\bL_1\bL_2}      & = -\tfrac{1}{g_-} \bigl( 2 D^{[\bL_1} A^{(1)\bL_2]}{}_{\alphaR} -  A^{(1)\aL}{}_{\alphaR} F_{\aL}{}^{\bL_1\bL_2} - A^{(1)\bL_1\betaR} A^{(1)\bL_2\gammaR}f_{\alphaR\betaR\gammaR} \bigr) \,.
\end{align}
Equations \eqref{eqn:A2a} and \eqref{eqn:A2alphaRbetaL} have their counterparts with conjugate chiralities. They follow the same pattern as for the previous orders, namely one has to send $g_-$ to $g_+$ and conjugate every index (exchanging over- and underlines). Therefore, we do not have to write them down explicitly. Finally, like in \eqref{eqn:defA1}, we define the index-free version
\begin{equation}\label{eqn:defA2}
	A^{(2)} = A^{(2)}_{a\beta} R^{a\beta} + \tfrac12 A^{(2)}_{\alpha\beta} R^{\beta\alpha}
\end{equation}
for later use.

At this point, we want to address the subtle point that the components $A_{\alphaL\betaR}$ and $A_{\alphaR\betaL}$ are not independent. Rather, they are related by
\begin{equation}
	A_{\alphaL\betaR} = - A_{\betaR\alphaL}\,.
\end{equation}
As a consequence, \eqref{eqn:lock2R} and \eqref{eqn:lock2L} have to be related and cannot be treated independently. Therefore, we have to check if \eqref{eqn:A2alphaRbetaL} and its conjugate satisfy
\begin{equation}\label{eqn:A2alphabetaantisym}
	S_{\alphaL}{}^{\gammaL} \left( A^{(2)}_{\gammaL\deltaR} + A^{(2)}_{\deltaR\gammaL} \right) S_{\betaR}{}^{\deltaR} = S_{\alphaL}{}^{\gammaL} \tA^{(2)}_{\gammaL\betaR} + \tA^{(2)}_{\deltaR\alphaL} S_{\betaR}{}^{\deltaR} = 0\,,
\end{equation}
where both sides have been multiplied by $S$ and $S^T$ to go from $A$'s to $\tA$'s. To also convert all the $A$'s in \eqref{eqn:A2alphaRbetaL} to $\tA$'s, we need to identify
\begin{equation}
	(S^{-1})_{\alpha_1}{}^{\delta} (S^{-1})_{\beta_1}{}^{\epsilon} S_{\gamma}{}^{\rho} f_{\delta\epsilon\rho} = f_{\alpha_1\beta_1\gamma_1}\,.
\end{equation}
Combining it with the identifications above, we find that \eqref{eqn:A2alphabetaantisym} is equivalent to
\begin{equation}
	D_{[\aL_1} F_{\aL_2\bR_1\bR_2]} - \tfrac34 F_{[\aL_1\aL_2}{}^C F_{\bR_1\bR_2]C} = 0\,,
\end{equation}
which vanishes due to the Bianchi identity
\begin{equation}
	D_{[A} F_{BCD]} - \tfrac34 F_{[AB}{}^E F_{CD]E} = 0
\end{equation}
for the generalized fluxes.

By now it should be clear how these computations proceed. We will do one more order because we need it to compute the gauge-fixing and with it the gGS transformations up to four derivatives in the next subsection. The relevant three-derivative contributions to the generalized fluxes are
\begin{equation}
	\begin{aligned}
		\bcF^{(3)}_A \cong & [ A^{(2)}, \bF_A ] + \tfrac12 [ A^{(1)}, [ A^{(1)}, \bF_A ] ] - \tfrac12 A^{(1)}_{A\gamma} A^{(1)B\gamma} \bF_B + \tfrac12 [A^{(1)}, D_A A^{(1)}] \\
		                   & - \tfrac12 A^{(1)}_{A\beta} [ A^{(1)}, [ A^{(1)}, R^{\beta} ] ] + \hlangle_A | D_B A^{(2)} \ZHet |^B \hrangle                                     \\
		                   & + \hlangle_A| D_B A^{(1)} \ZHet A^{(1)} |^B \hrangle + \tfrac12 \hlangle_A| [ A^{(1)}, D_B A^{(1)}] \ZHet |^B \hrangle\,.
	\end{aligned}
\end{equation}
Moreover, we now need to take $\bcG^{(3)}_{\Ac}$ into account. It contributes with
\begin{equation}
	\bcG^{(3)\beta}_{A} \cong - \left(c_1 + \tfrac16\right) A^{(1)}_A{}^\beta A^{(1)}_{B\beta} A^{(1) B}_\gamma t^\gamma\,,
	\qquad \text{and} \qquad
	\bcG^{(3)\beta}_{\alpha} = 0\,.
\end{equation}
In particular, we use the only non-vanishing component
\begin{equation}
	\tbcG^{(3)}_{\aR\betaL} \cong - \left(c_1 + \tfrac16\right)  A^{(1)\gammaL}_{\aR} A^{(1)}_{\bR\gammaL} \tA^{(1)\bR}_{\betaL}
\end{equation}
to compute
\begin{align} \label{eqn:A31}
	K:   &  & \tA_{\aR}^{(3)\bL_1\bL_2}        & =  \tfrac{1}{g_-} \Bigl( A_{\gammaR}^{(1)[\bL_1} D_{\aR} A^{(1)\bL_2]\gammaR} -A_{\gammaR}^{(1)[\bL_1|} A^{(1)}_{\cL}{}^{\gammaR} F_{\aR}{}^{|\bL_2]\cL} \Bigr)\,,                                             \\
	\intertext{and furthermore}
	R_1: &  & \tA^{(3)}_{\aR}{}^{\bL}_{\betaL} & = \tfrac{1}{2 g_-} \left( A^{(2)}_{\cR\betaL} F_{\aR}{}^{\bL\cR} + D^{\bL} A^{(2)}_{\aR\betaL} \right)\label{eqn:A33}\,,                                                                                       \\
	R_2: &  & \tA^{(3)}_{\aR\betaL\gammaL}     & = \tfrac{1}{g_-} \Bigl( A^{(1)}_{\bR[\betaL|} D_{\aR} A^{(1) \bR}_{|\gammaL]} - A^{(1) \bR}_{[\betaL} A^{(1) \cR}_{\gammaL]} F_{\aR\bR\cR} + 2 D_{\bR} A^{(1)}_{\aR[\betaL} A^{(1) \bR}_{\gammaL]} + \nonumber \\ & & &
	\qquad\qquad\qquad\qquad A^{(1)\alphaL}_{\aR} f_{\deltaL[\betaL|\alphaL} A^{(1)}_{\bR|\gammaL]} A^{(1) \bR \deltaL} \Bigr)\,,
\end{align}
which both do not receive corrections from $\bcG^{(3)}_{\aR}$. At this point something remarkable happens: For a general $c_1$ in the parameterization \eqref{eqn:Aprime}, one finds a third term in \eqref{eqn:A31}. However, tuning it to
\begin{equation}
	c_1 = \frac{1}{3}
\end{equation}
this term vanishes. In the next section, we will see that this choice also leads to significant simplifications of the gauge transformations. There, we will adopt it here. $\bcF^{(3)}$ has one more component,
\begin{equation}
	\begin{aligned}
		\bcF^{(3)\alpha} \cong & \left(A^{(1)B \alpha} A^{(1)}_C{}^{\beta}F_{B A}{}^C +\tfrac{1}{2}f^{\gamma\delta\alpha}A^{(1)}_{A\gamma}A^{(2)\beta}_{\delta} + D_A A^{(2)\beta\alpha} + D_B A^{(1)}_A{}^{\alpha}A^{(1)B\beta}\right)R^A_{\beta} \\
		                       & +\left(-A^{(2)A\alpha}F_{A B C} + f^{\beta\gamma\alpha}A^{(1)}_{C\beta}A^{(1)}_{B\gamma} + 2 D_B A^{(2)}_C{}^{\alpha}\right)K^{B C}\,,
	\end{aligned}
\end{equation}
which gives rise to
\begin{align}
	K:   &  & \tA^{(3)\alphaR\bL_1\bL_2}      & = \tfrac{1}{g_-}\left( A^{(2)\aL\alphaR} F_{\aL}{}^{\bL_1\bL_2} - f^{\deltaR\gammaR\alphaR} A^{(1)[\bL_1}{}_{\deltaR} A^{(2)\bL_2]}{}_{\gammaR} + 2 D^{[\bL_1} A^{(2)\bL_2]\alphaR} \right),                                                                                                   \\
	R_1: &  & \tA^{(3)\alphaR \bL}{}_{\betaL} & = -\tfrac{1}{g_-} \Bigl( A^{(1)\dL\alphaR} A^{(1)}_{\cR\betaL} F_{\dL}{}^{\bL\cR}+\tfrac{1}{2} f_{\gammaR\deltaR}{}^{\alphaR} A^{(1)\bL\gammaR} A^{(2)\deltaR}{}_{\betaL} +                    D^{\bL} A^{(2)}_{\betaL}{}^{\alphaR} + D_{\dR} A^{(1)\bL\alphaR} A^{(1)\dR}{}_{\betaL}\Bigr)\,.
\end{align}
Again, we combine all these components into
\begin{equation}\label{eqn:defA3}
	A^{(3)} = A^{(3)}_{a\beta} R^{a\beta} + \tfrac12 A^{(3)}_{\alpha\beta} R^{\beta\alpha}\,.
\end{equation}

\subsection{Generalized Green-Schwarz transformations}\label{sec:genGSTr}
In the last subsection, we have seen how the identification of the connection $\cA$ proceeds order by order. Now, we will revisit the gauge transformations from section~\ref{sec:gaugeTr} and treat them in the same way. Not only this will confirm that the gauge fixing we proposed in section~\ref{sec:gaugeFixing} is valid, but it will also give rise to the gGS transformations of the physical frame. Knowing them, will allow us to compare our results with the literature on the gBdR identification. Like before, we will expand $\delta_\xi \mathbb{E}$ according to the number of derivatives it carries. Formally, this will always look like
\begin{equation}
	\delta \mathbb{E}^{(l)} = \delta A'^{(l)} + \delta E^{(l)} E^{-1} + \dots = \xi^{(l)\alpha} t_\alpha + \dots\,,
\end{equation}
where $\dots$ denotes lower-order contributions. We bring them all to one side of the equation and denote them by $\cX^{(l)}[\xi^{(<l)}]$. As before, $\xi^{(<l)}$ indicates contributions of lower order, with the highest being $\xi^{(l-1)}$. In this way, there is a clear distinction between already-known and yet-to-fix quantities,
\begin{equation}\label{eqn:gaugeIdentStep1}
	\boxed{%
		\delta A^{(l)} + \delta E^{(l)} E^{-1} - \xi^{(l)\alpha} t_\alpha = \cX^{(l)}[\xi^{(<l)}] - \delta G^{(l)}[\xi^{(<l)}]}
\end{equation}
with
\begin{equation}
	\cX = - D_A \xi^\beta R_\beta^A + \sum_{m=1}^{\infty} \frac{(-1)^m}{m!} \left[ A' ,\xi^\alpha \htau_\alpha - \tfrac{1}{m+1} \delta A' \right]_m\,,
\end{equation}
and
\begin{equation}
	\delta G^{(l)}[\xi^{(<l)}] = \delta A'^{(l)} - \delta A^{(l)}\,.
\end{equation}
Similar to the last subsection, we find for example
\begin{align}
	\delta G^{(1)} & = 0 \,,                                    \\
	\delta G^{(2)} & = 0 \,,
	\intertext{with the first non-trivial contribution being}
	\delta G^{(3)} & = c_1 \delta \left[ (A^{(1)})^3 \right]\,.
\end{align}
As in \eqref{eqn:identStep1}, we find on the left-hand-side of \eqref{eqn:gaugeIdentStep1} $t_\alpha$ instead of $\tau_\alpha$. Thus, we have to proceed in the same way as for the identification in the last subsection by relating
\begin{equation}
	\xi^{(l)\alpha} t_\alpha  = \txi^{(l)\alpha} \htau_\alpha
\end{equation}
in analogy with \eqref{eqn:AtildeToA}, or equally
\begin{equation}
	\xi^{(l)\alpha} = \txi^{(l)\beta} (S^{-1})_{\beta}{}^{\alpha}\,.
\end{equation}
Our gauge fixing requires that all chiral and anti-chiral contributions to $\cX^{(l)}$ for $i\ge1$ originate from the third term on the right-hand side. Hence, it comes down to a similar identification we already performed to fix $\tA^{(i)}$, namely
\begin{equation}\label{eqn:identxi}
	\begin{aligned}
		\txi^{(l)\alphaL} & = - \frac{1}{g_-^2} \llangle \cX^{(l)}[\xi^{(<l)}] - \delta G^{(l)}[\xi^{(<l)}], \htau^{\alphaL} \rrangle\,,  \\
		\txi^{(l)\alphaR} & = - \frac{1}{g_+^2} \llangle \cX^{(l)}[\xi^{(<l)}] - \delta G^{(l)}[\xi^{(<l)}], \htau^{\alphaR} \rrangle \,.
	\end{aligned}
\end{equation}

In evaluating $\cX^{(l)}$ explicitly, it is helpful to introduce some auxiliary quantities. Very often, we will deal with the contractions $\xi^\alpha t_\alpha$ and $\tilde{\xi}^\alpha \htau_\alpha$. Therefore, they are assigned to the corresponding expansions
\begin{equation}\label{eqn:xi}
	\txi = \txi^\alpha \htau_\alpha = \xi^\alpha t_\alpha = \Xi^{A B} K_{AB} + \Xi^\beta_A R_\beta^A + \tfrac12 \Xi^{\alpha\beta} R_{\beta\alpha}\,,
\end{equation}
or equally,
\begin{equation}\label{eqn:zeta}
	\xi = \xi^\alpha \htau_\alpha = \txi + \xi^\alpha R_\alpha\,.
\end{equation}
Starting at the leading order, we have
\begin{equation}\label{eqn:deltaE0}
	\delta E^{(0)} E^{-1} = \txi^{(0)}
\end{equation}
and therefore fix
\begin{equation}\label{eqn:xi0}
	\txi^{(0)} = - \Lambda^{\aL_1\aL_2} K_{\aL_1\aL_2} - \Lambda^{\aR_1\aR_2} K_{\aR_1\aR_2}\,,
\end{equation}
allowing us to read off
\begin{equation}
	\Xi^{(0)\aL_1\aL_2} = - \Lambda^{\aL_1\aL_2} \,, \qquad \text{and} \qquad
	\Xi^{(0)\aR_1\aR_2} = - \Lambda^{\aR_1\aR_2}\,.
\end{equation}
There are several things to explain here. Most important is that the form of \eqref{eqn:xi0} is motivated by standard double Lorentz transformations with the parameters $\Lambda^{\aL\bL}$ and $\Lambda^{\aR\bR}$ (both of them are anti-symmetric in their two indices). Our sign choice is dictated by the conventions used in \cite{Baron:2020xel} because we later want to match their results for the gGS transformations. Taking into account \eqref{eqn:xi} and the generators \eqref{eqn:tau00}, we find
\begin{equation}\label{eqn:xi0alpha}
	\txi^{(0)\aL_1\aL_2} = - \tfrac{1}{g_-}\Lambda^{\aL_1\aL_2} \,,\qquad \text{and} \qquad
	\txi^{(0)\aR_1\aR_2} = - \tfrac{1}{g_+}\Lambda^{\aR_1\aR_2}
\end{equation}
as the only non-vanishing contributions to $\txi^{(0)}$. Moreover, we will perform the gauge fixing such that this is the only contribution to $\delta E_{\aL\bL}$ and $\delta E_{\aR\bR}$. Therefore,
\begin{equation}
	\delta E_{\aL\bL} = - \Lambda_{\aL\bL} \,, \qquad \text{and} \qquad
	\delta E_{\aR\bR} = - \Lambda_{\aR\bR}\,,
\end{equation}
with $\delta E_{ab} = \hlangle_a | \delta E E^{-1} |_b \hrangle$ holding to all orders, and we have fixed
\begin{equation}
	\xi^{(0)} = \Xi^{(0)AB} K_{AB} + \xi^{(0)\gamma} R_{\gamma}\,.
\end{equation}
Moreover, note that $E$ as the fundamental field does not receive higher-derivative corrections -- in contrast to its gauge transformations $\delta E$. Thus, we write only $E^{-1}$ in \eqref{eqn:deltaE0}.

At the next order, we encounter
\begin{equation}
	\cX^{(1)} = - D_A \xi^{(0)\beta} R_\beta^A - [ A^{(1)}, \xi^{(0)} ],
\end{equation}
with $A^{(1)}$ from \eqref{eqn:defA1}. One notices that writing out all the different projections on the first term is cumbersome. Therefore, we introduce the shorthand notation
\begin{equation}
	D \xi^{(l)}_+ = D_{\aL} \xi^{(l)\betaL} R_{\betaL}^{\aL} + D_{\aR} \xi^{(l)\betaR} R_{\betaR}^{\aR}
\end{equation}
for chiral/anti-chiral contributions and
\begin{equation}
	D \xi^{(l)}_- = D_{\aL} \xi^{(l)\betaR} R_{\betaR}^{\aL} + D_{\aR} \xi^{(l)\betaL} R_{\betaL}^{\aR}
\end{equation}
for mixed chirality generators, to write
\begin{equation}
	\cX^{(1)} = - D \xi^{(0)}_+ - D \xi^{(0)}_- - [A^{(1)}, \xi^{(0)}]\,.
\end{equation}
In this form, one can already do the identification at the level of \eqref{eqn:gaugeIdentStep1} with the result
\begin{align}\label{eqn:deltaA1}
	\delta A^{(1)} & = - D \xi^{(0)}_- - [ A^{(1)}, \xi^{(0)} ]\,, \qquad \text{and} \\ \label{eqn:defxi1}
	\txi^{(1)}     & = D \xi^{(0)}_+\,.
\end{align}
From the first equation, we get the transformation
\begin{equation}
	\delta \tA^{(1)}_{\aR \betaL_1} = -D_{\aR} \txi^{(0)}_{\betaL_1} + \Xi^{(0)\cR}_{\aR} \tA^{(1)}_{\cR\betaL_1} + \tA^{(1)\gammaL_1}_{\aR}\txi^{(0)\deltaL_1}f_{\gammaL_1\deltaL_1\betaL_1}\, ,
\end{equation}
where we used
\begin{equation}
	(S^{-1})_{\alpha_1}{}^{\gamma}(S^{-1})_{\beta_1}{}^{\delta}f_{\gamma\delta}{}^{\epsilon} = f_{\alpha_1\beta_1}{}^{\gamma_1}(S^{-1})_{\gamma_1}{}^{\epsilon}
\end{equation}
to transition to $\tA$. It originates directly from \eqref{eqn:SinvSinvf}. Alternatively, one can compute it by using the result for $\tA^{(1)}$ obtained in \eqref{eqn:identA1} and the leading order transformation of the physical frame. Both match and thereby provide a consistency check. The same holds for the conjugate chirality. From equation \eqref{eqn:defxi1}, we also get
\begin{align}
	R_1: \qquad\qquad \Xi^{(1)\alphaL}_{\aL} & = D_{\aL}\xi^{(0)\alphaL}                 \\
	\intertext{resulting in}
	\txi^{(1)\alphaL}_{\aL}                  & =\tfrac{2}{g_-}\Xi^{(1)\alphaL}_{\aL} \,,
\end{align}
as non-vanishing contributions to $\txi^{(1)\alpha}$.  In conclusion, we have performed the first step of the gauge fixing, resulting in
\begin{equation}
	\xi^{(1)} = D \xi^{(0)}_+ + \xi^{(1)\alpha} R_\alpha\,.
\end{equation}

\subsubsection{Leading order}
The next order is already more complicated with
\begin{equation}
	\cX^{(2)} = - D \xi^{(1)}_+ - D \xi^{(1)}_- - [ A^{(1)}, \xi^{(1)} ] - [ A^{(2)}, \xi^{(0)} ] + \tfrac12 [ A^{(1)}, [A^{(1)}, \xi^{(0)}] ] + \tfrac12 [ A^{(1)}, \delta A^{(1)} ]\,,
\end{equation}
after making use of the index-free $A^{(2)}$ defined in \eqref{eqn:defA2}. It decomposes into a (anti-)chiral and mixed chirality part, which are used to fix
\begin{align}\label{eqn:deltaA2}
	\delta A^{(2)} + \delta E^{(2)} E^{-1} & = - D \xi^{(1)}_- - [ A^{(1)}, \xi^{(1)} ] - [ A^{(2)}, \xi^{(0)} ] \,, \qquad \text{and}                        \\
	\xi^{(2)}                              & = D \xi^{(1)}_+ - \tfrac12 [ A^{(1)}, [ A^{(1)} , \xi^{(0)} ] ] - \tfrac12 [ A^{(1)}, \delta A^{(1)} ] \nonumber \\
	                                       & = D \xi^{(1)}_+ + \tfrac12 [ A^{(1)}, D \xi^{(0)}_- ]
\end{align}
by comparing with \eqref{eqn:gaugeIdentStep1}.
Here, we obtain the last line by using \eqref{eqn:deltaA1}. In order to isolate $\delta A^{(2)}$ from $\delta E^{(2)} E^{-1}$, we have to project the first equation onto the generators $K^{AB}$ to get
\begin{equation}\label{eqn:deltaE2pre}
	\delta E^{(2)}_{\aL\bR} = - \hlangle_{\aL} | [ A^{(1)}, \xi^{(1)}] |_{\bR}\hrangle =  A^{(1)}_{\aL\alphaR}D_{\bR}\xi^{(0)\alphaR} -  A^{(1)}_{\bR\alphaL} D_{\aL} \xi^{(0)\alphaL} = A^{(1)}_{\aL\alphaR}D_{\bR}\xi^{(0)\alphaR} - \text{c.c.}\,.
\end{equation}
For convenience, we introduced the operation $\text{c.c.}$ which abbreviates conjugate chirality. It acts on each term by flipping $\aL \leftrightarrow \bR$ and the chiralities of all dummy indices. Finally, we rewrite the left-hand side in terms of tilded quantities that we already computed. This is the first time, we have to collapse two towers along the lines of section~\ref{sec:collapsingTowers}. We will do it therefore in more detail by writing the first term on the right-hand-side of \eqref{eqn:deltaE2pre} as
\begin{equation}
	A^{(1)\alphaR}_{\aL} D_{\bR}\xi^{(0)\betaR} \kappa_{\alphaR\betaR} =
	\tA^{(1)\alphaR}_{\gammaL} D_{\bR}\txi^{(0)\deltaR} \tkappa_{\alphaR\betaR} = \chiR \tA^{(1)}_{\aL\alphaR_1} D_{\bR}\txi^{(0)\alphaR_1}
\end{equation}
to obtain
\begin{equation}
	\delta E^{(2)}_{\aL\bR} = \chiR \tA^{(1)}_{\aL\alphaR_1} D_{\bR}\txi^{(0)\alphaR_1}  - \text{c.c.} \,.
\end{equation}
Plugging in the explicit expressions \eqref{eqn:identA1} for $\tA^{(1)}_{\aL\alphaR_1}$ and \eqref{eqn:xi0alpha} for $\txi^{(0)\alphaR_1}$ presented above, we get the leading order GS transformation
\begin{equation}
	\delta E^{(2)}_{\aL\bR} = - \frac{\chiL}{g_-^2} D_{\aL} \Lambda^{\cL_1\cL_2} F_{\bR\cL_1\cL_2} + \frac{\chiR}{g_+^2} D_{\bR} \Lambda^{\cR_1\cR_2} F_{\aL\cR_1\cR_2} =
	\frac{a}{2} D_{\aL} \Lambda^{\cL_1\cL_2} F_{\bR\cL_1\cL_2} + \frac{b}{2} D_{\bR} \Lambda^{\cR_1\cR_2} F_{\aL\cR_1\cR_2} \,.
\end{equation}
Here, the right-hand side is the result for the leading order in \cite{Baron:2020xel}. Therefore, we can finally fix
\begin{equation}\label{eqn:valuesconstLR}
	\boxed{%
		a = -\frac{2\chiL}{g_-^2}\,, \qquad \text{and} \qquad b = \frac{2\chiR}{g_+^2}\,.}
\end{equation}
At the same time, we also extract the gauge-fixing constraints
\begin{align}
	K: \qquad  \Xi^{(2)}_{\aL\bL}          & = - A^{(1)}_{[\aL|\betaR} D_{|\bL]} \xi^{(0)\betaR} \,,     \\
	R_1: \qquad  \Xi^{(2)}_{\aL\betaL}     & = \phantom{-} D_{\aL} \xi^{(1)}_{\betaL} \,,                \\
	R_2: \qquad  \Xi^{(2)}_{\alphaL\betaL} & = - A^{(1)\cR}_{[\alphaL|} D_{\cR} \xi^{(0)}_{|\betaL]} \,,
\end{align}
which complete
\begin{equation}
	\xi^{(2)} = \Xi^{(2)}_{AB} K^{AB} + \Xi^{(2)\beta}_A R_{\beta}^A - \tfrac12 \Xi^{(2)\alpha\beta} R_{\alpha\beta} + \xi^{(2)\alpha} R_\alpha\,.
\end{equation}
In particular, we will need the components
\begin{align}
	\txi^{(2)}_{\aL_1\aL_2}    & = -\tfrac{1}{g_-} A^{(1)}_{[\aL_1 | \betaR} D_{|\aL_2]} \xi^{(0)\betaR}\,, \qquad \text{and} \qquad \\
	\txi^{(2)}_{\alphaL\betaL} & = -\tfrac{1}{g_-} A^{(1)\cR}_{[\alphaL|} D_{\cR} \xi^{(0)}_{|\betaL]}
\end{align}
for the gGS transformations with four derivatives. This process continues to higher orders, but we will not pursue it in full detail here. The only exception is
\begin{equation}\label{eqn:xi3}
	\txi^{(3)} = D\xi^{(2)}_+ + \tfrac12 [ A^{(2)}, D\xi^{(0)}_- ] + \tfrac12 [ A^{(1)}, D\xi_-^{(1)} + \delta E^{(2)} E^{-1} ]
\end{equation}
which is needed later. Note that here $\delta G^{(3)}$ does not give additional contributions.

\subsubsection{Next to leading order}
To evaluate $\delta E^{(4)}_{\aL\bR}$, we have to work with
\begin{equation}
	\cX^{(4)}_- \cong - [A' , \xi ] - \tfrac16 [ A, \xi ]_3 - \tfrac16 [ A, \delta A ]_2 \Bigr|^{(4)} \,,
\end{equation}
where the left-hand side is restricted to terms with only four derivatives. Remarkably, this expression can be simplified considerably by using the results for $\delta A^{(1)}$ and $\delta A^{(2)}$ we obtained above. After some rearrangement, we find
\begin{equation}
	\cX^{(4)}_- \cong - [ A^{(1)}, \dxi^{(3)} ] - [ A^{(2)}, \dxi^{(2)} ] - [ A'^{(3)}, \txi^{(1)} ] - [ A'^{(4)}, \txi^{(0)} ],
\end{equation}
with
\begin{equation}
	\dxi^{(l)} = \xi^{(l)} - \tfrac13 \left( \txi^{(l)} - D\xi^{(l-1)}_+ \right)\,,
\end{equation}
where we have used \eqref{eqn:deltaA1},\eqref{eqn:deltaA2} and \eqref{eqn:xi3}. From \eqref{eqn:Aprime}, we immediately see
\begin{align}
	A'^{(3)} & = A^{(3)} + c_1 (A^{(1)})^3  \,, \qquad \text{and}                                           \\
	A'^{(4)} & = c_1 \left( A^{(2)} (A^{(1)})^2 + A^{(1)} A^{(2)} A^{(1)} + (A^{(1)})^2 A^{(2)} \right) \,.
\end{align}
Up to $R^{\alpha\beta}$ generators, which will not contribute to $\delta E^{(4)} E^{-1}$ (denoted again by $\cong$), we moreover have
\begin{align}
	\dxi^{(2)} & \cong D \xi^{(1)}_+ \,, \qquad \text{and}                                                                                \\
	\dxi^{(3)} & \cong D \xi^{(2)}_+ + \tfrac13 [ A^{(2)}, D\xi^{(0)}_- ] + \tfrac13 [ A^{(1)}, D\xi^{(1)}_- + \delta E^{(2)} E^{-1} ]\,.
\end{align}
At this point, we also need to compute $\delta G^{(4)}$ because it now contributes to the final result. One can actually show that
\begin{align}
	\delta G^{(4)}_- & = c_1 \delta \left[ A^{(2)} A^{(1)} A^{(1)} + A^{(1)} A^{(2)} A^{(1)} + A^{(1)} A^{(1)} A^{(2)} \right] \\
	                 & \cong
	- 3 c_1 [ A^{(1)}, \dxi^{(3)}] + 3 c_1 [ A^{(1)}, D\xi^{(2)}_+ ] + [ A^{(3)}, \txi^{(1)} ] - [ A'^{(3)}, \txi^{(1)} ] - [ A'^{(4)}, \txi^{(0)} ]
\end{align}
holds. By now it becomes obvious again that $c_1 = 1/3$ is the preferred choice for which the transformation simplifies considerably and we are left with
\begin{equation}
	\delta E^{(4)} E^{-1} \cong \cX^{(4)}_- - \delta G^{(4)} \cong - [A^{(1)}, D \xi_+^{(2)} ] - [A^{(2)}, D \xi_+^{(1)} ] - [A^{(3)}, D \xi_+^{(0)} ]\,.
\end{equation}
In indices, this relation results in the gGS transformation
\begin{equation}\label{eqn:deltaE4}
	\delta E^{(4)}_{\aL\bR}  = A^{(3)}_{\aL\alphaR}D_{\bR}\xi^{(0)\alphaR}+ A^{(2)}_{\aL\alphaR}D_{\bR}\xi^{(1)\alphaR} +A^{(1)}_{\aL\alphaR}D_{\bR}\xi^{(2)\alphaR} - \text{c.c.}\,.
\end{equation}
This is a very nice result that allows us to write the universal form
\begin{equation}\label{eqn:gGSmaster}
	\boxed{%
	\delta E_{\aL\bR}^{(2 m)} = \sum_{n=1}^{2m-1} A^{(n)}_{\aL \alphaR} \, D_{\bR} \xi^{(2n-1-i)\alphaR} - \text{c.c} = \left. A_{\aL\alphaR} D_{\bR} \xi^{\alphaR} - \text{c.c.}\right|^{(2 m)}}
\end{equation}
of gGS transformations up to $m\le2$. We plan to investigate if there are appropriate choices for $c_n$, $n>1$ such that this relation continues to hold at higher orders.

Like in the leading order, the last step is to collapse all the towers and thereby go from untilded $A$'s and $\xi$'s to tilded ones. We do not have a choice because we only have explicit expressions for the latter. However, now the process is more complicated because we have to do it twice. The first iteration is straightforward and gives rise to
\begin{equation}\label{eqn:deltaE4collapse1}
	\begin{aligned}
		\delta E_{\aL\bR} = & \chiR \tA^{(3)}_{\aL\alphaR_1} D_{\bR} \txi^{(0)\alphaR_1} + \chiR \tA^{(2)}_{\aL\alphaR} D_{\bR} \txi^{(1)\alphaR} + \chiR \tA^{(1)}_{\aL\alphaR_1} D_{\bR}\txi^{(2)\alphaR_1} +                     \\
		                    & \chiR_1 \tA^{(3)}_{\aL\alphaR} D_{\bR} \txi^{(0)\betaR_1}(\tSinv)_{\betaR_1}{}^{\alphaR} + \chiR \tA^{(1)\alphaR_1}_{\aL}D_{\bR} \txi^{(2)}_{\betaR}(\tSinv)_{\alphaR_1}{}^{\betaR} - \text{c.c.} \,.
	\end{aligned}
\end{equation}
Next, we go over each term on the right-hand side. Take for example the first one: There, we have to collapse a second set of towers inside \eqref{eqn:A31} for $\tA_{\aR\betaR_1}$ giving rise to
\begin{equation}
	\tA^{(3)}_{\aL\bR_1\bR_2} = \tfrac{\chiR}{g_+} \Bigl( \tA^{(1)}_{[\bR_1|\gammaL_1} D_{\aL} \tA^{(1)}_{|\bR_2]}{}^{\gammaL_1} - \tA^{(1)}_{[\bR_1|\gammaL_1} \tA^{(1)}_{\cR}{}^{\gammaL_1} F_{\aL|\bR_2]}{}^{\cR} \Bigr)\,.
\end{equation}
The remaining terms on the first line of \eqref{eqn:deltaE4collapse1} follow the same pattern. We can easily transition to tilded quantities by just to adding a global $\chiR$ factor. For a few terms on the second line, in addition to $\tkappa_{\alpha\beta}$ also knowledge of $\tf_{\alphaL_1\betaL_1\gammaL_1}$ and $\tf_{\alphaL_1\betaL_1\epsilonL}\tf_{\gammaL_1\deltaL_1\rhoL}\tkappa^{\epsilonL\rhoL}$ are required. Fortunately, we have already computed them at the end of subsection~\ref{sec:collapsingTowers}. After writing everything out in terms of the generalized fluxes, double Lorentz parameters and their flat derivatives, we eventually are left with
\begin{equation}\label{eqn:deltaE4final}
	\begin{aligned}
		\delta E^{(4)}_{\aL\bR} = - \frac{a^2 }{2} & \left[D_{\aL}D_{\cL}\Lambda_{\dL\eL}\left(F^{\cL}{}_{\fR\bR}F^{\fR\dL\eL}+D^{\cL}F_{\bR}{}^{\dL\eL}\right) - F_{\bR\fL}{}^{\gL}F^{\cR}{}_{\dL\gL}\left(F_{\cR}{}^{\eL\dL}D_{\aL}\Lambda_{\eL}{}^{\fL}-F_{\cR}{}^{\eL\fL}D_{\aL}\Lambda_{\eL}{}^{\dL}\right) \right. \\
		                                           & \left. + D_{\aL}\Lambda_{\eL\fL}F^{\cR\eL}{}_{\dL}\left(F_{\bR\cR\gR}F^{\gR\fL\dL}-D_{\bR}F_{\cR}{}^{\fL\dL}+2D_{\cR}F_{\bR}{}^{\fL\dL}\right)+F_{\bR}{}^{\eL\dL}D_{\aL}\left(D^{\cR}\Lambda_{\eL}{}^{\fL}F_{\cR\fL\dL}\right)\right]                               \\
		- \frac{a b}{4}                            & \left[D_{\aL}\Lambda^{\cL\dL}\left(F_{\bR\cL\gL}F^{\gL\eR\fR}F_{\dL\eR\fR}-D_{\bR}F_{\cL}{}^{\eR\fR}F_{\dL\eR\fR}\right)+F_{\bR\cL\dL}D_{\aL}\left(D^{\cL}\Lambda_{\eR\fR}F^{\dL\eR\fR}\right)\right.                                                               \\
		                                           & \left. - D_{\bR}\Lambda^{\cR\dR}\left(F_{\aL\cR\gR}F^{\gR\eL\fL}F_{\dR\eL\fL}-D_{\aL}F_{\cR}{}^{\eL\fL}F_{\dR\eL\fL}\right)-F_{\aL\cR\dR}D_{\bR}\left(D^{\cR}\Lambda_{\eL\fL}F^{\dR\eL\fL}\right)\right]                                                            \\
		+ \frac{b^2}{2}                            & \left[ D_{\bR}D_{\cR}\Lambda_{\dR\eR}\left(F^{\cR}{}_{\fL\aL}F^{\fL\dR\eR}+D^{\cR}F_{\aL}{}^{\dR\eR}\right) -F_{\aL\fR}{}^{\gR}F^{\cL}{}_{\dR\gR}\left(F_{\cL}{}^{\eR\dR}D_{\bR}\Lambda_{\eR}{}^{\fR}-F_{\cL}{}^{\eR\fR}D_{\bR}\Lambda_{\eR}{}^{\dR}\right)\right.  \\
		                                           & \left. + D_{\bR}\Lambda_{\eR\fR}F^{\cL\eR}{}_{\dR}\left(F_{\aL\cL\gL}F^{\gL\fR\dR}-D_{\aL}F_{\cL}{}^{\fR\dR} + 2 D_{\cL}F_{\aL}{}^{\fR\dR} \right) +F_{\aL}{}^{\eR\dR}D_{\bR}\left(D^{\cL}\Lambda_{\eR}{}^{\fR}F_{\cL\fR\dR}\right) \right]\,.
	\end{aligned}
\end{equation}
As for the leading order expression, we have substituted the values for $a$ and $b$ obtained in \eqref{eqn:valuesconstLR}. Our result perfectly matches the result for the bi-parametric deformation presented in \cite{Baron:2020xel}.

At this point, we see that the simple transformation \eqref{eqn:gGSmaster} becomes very complicated after collapsing all towers and using the results of the connection $\cA$ from the last subsection. A drawback of the original gBdR identification is that it does not give access to the intermediate results we have used to finally get to \eqref{eqn:deltaE4final}; it is all or nothing. Besides the additional computational complexibility this implies, without being able to reuse intermediate results, one always has to start from scratch for each new order -- this obfuscates important structures like \eqref{eqn:gGSmaster}.

\subsection{Action on the mega-space}\label{sec:PSaction}
A perk of the \PS{} construction is that it allows to easily construct invariant quantities under the symmetry it implements. Such quantities are essential to construct actions for physical theories. In general, there is more than one invariant. Hence, which one should we choose for the action? Our guiding principle for this equation is that we know that at leading order, it should be the standard double field theory action. Only in higher orders, it will receive corrections which are then completely fixed by symmetry. The same idea is used in the  original gBdR identification. There is a natural candidate for such an action, namely the two-derivative action on the mega-space \cite{HasslerHetPS:2024}. It can be written in a manifest left-right symmetric form as
\begin{equation}
	S = \int \dd^d x e^{-2 \Phi} \Rc\, \qquad \text{with} \qquad
	\Rc = \Rc_0 + \Rc_1 + \Rc_2\,,
\end{equation}
where $\Rc_0$ is a constant, $\Rc_1$ contains the dilatonic reduced twisted torsion $\THet_{\Ac}$ through
\begin{equation}\label{eqn:R1}
	\Rc_1 = \Bigl( 2 \nabla^{\AhL} \THet_{\AhL} - \THet^{\AhL} \THet_{\AhL} \Bigr) - \Bigl( 2 \nabla^{\AhR} \THet_{\AhR} - \THet^{\AhR} \THet_{\AhR} \Bigr)\,,
\end{equation}
and $\Rc_2$ accommodates the reduced twisted torsion $\THet_{\Ac\Bc\Cc}$ as
\begin{equation}\label{eqn:R2}
	\Rc_2 = \Bigl( \tfrac12 \THet_{\AhR\BhL\ChL} \THet^{\AhR\BhL\ChL} + \tfrac16 \THet_{\AhL\BhL\ChL} \THet^{\AhL\BhL\ChL} \Bigr) - \left( \tfrac12 \THet_{\AhL\BhR\ChR} \THet^{\AhL\BhR\ChR} + \tfrac16 \THet_{\AhR\BhR\ChR} \THet^{\AhR\BhR\ChR} \right)\,.
\end{equation}
It has a $\mathbb{Z}^2$ symmetry under the exchange of chiral and anti-chiral projectors and flipping the sign in front of the action. There are some new objects appearing here that we need to explain. Most important are the full heterotic indices with the pairing
\begin{equation}
	\heta_{\Ah\Bh} = \begin{pmatrix}
		- \kappa_{\alpha\beta} & 0              \\
		0                      & \heta_{\Ac\Bc}
	\end{pmatrix}\,,
	\qquad \text{and} \qquad
	\heta^{\Ah\Bh} = \begin{pmatrix}
		- \kappa^{\alpha\beta} & 0              \\
		0                      & \heta^{\Ac\Bc}
	\end{pmatrix}\,.
\end{equation}
They have not played any role yet, because all physically relevant information is already encoded in $\THet_{\Ac}$ and $\THet_{\Ac\Bc\Cc}$. However, only the action with the full indices is invariant as has been discussed in \cite{HasslerHetPS:2024}. Over- and underbars on these indices are just pulled through to their individual components. As the structure constants of our structure group satisfy $f_{\alpha\beta}{}^\beta = 0$, we find that $\THet_\alpha = 0$ and therewith
\begin{equation}
	\THet_{\Ah} = \begin{pmatrix}
		0 \quad & \THet_{\Ac}
	\end{pmatrix}\,.
\end{equation}
Combining it with the covariant derivative $\nabla^{\Ah}$ \cite{HasslerHetPS:2024} results in
\begin{equation}
	\nabla_{\Ah} \THet^{\Ah} = \cD_{\Ac} \THet^{\Ac} + \cA_{\Ac\gamma} (\htau^{\gamma})^{\Ac\Bc} \THet_{\Bc} := \nabla_{\Ac} \THet^{\Ac}\,.
\end{equation}
This is all we need to rewrite $\Rc_1$ from \eqref{eqn:R1} exclusively in terms of quantities we have already computed. The same is possible for $\Rc_2$ given by \eqref{eqn:R2}. However, the situation here is more subtle because in contrast to $\THet_{\alpha}=0$ the component $\THet_{\alpha\Bh\Ch}$ does not vanish. But fortunately, it is constant and thus any contraction in $\Rc_2$ involving a lowered Greek index will be constant and can be shifted to $\Rc_0$. By ignoring all constant contributions, we define
\begin{equation}\label{eqn:Rs}
	\begin{aligned}
		\Rc_1 & \cong 2 \nabla_{\AcR} \THet^{\AcR} - \THet_{\AcR} \THet^{\AcR} - \text{c.c.}\,,                                                 \\
		\Rc_2 & \cong \tfrac12 \THet_{\AcL\BcR\CcR} \THet^{\AcL\BcR\CcR} + \tfrac16 \THet_{\AcR\BcR\CcR} \THet^{\AcR\BcR\CcR} - \text{c.c.} \,.
	\end{aligned}
\end{equation}
Here $\simeq$ denotes up to constant terms that we always remove by a proper choice of $\Rc_0$. To make direct contact with the original gBdR identification, it is preferable to rewrite the action in terms of the heterotic fluxes $\cF_{\Ac\Bc\Cc}$, $\cF_{\Ac}$ instead of the reduced twisted torsions $\THet_{\Ac\Bc\Cc}$, $\THet_{\Ac}$. As the first step in this direction, note that $\THet_{\Ac\Bc\Cc}$ still contains the constant contribution
\begin{equation}\label{eqn:THet0}
	\THet^{(0)}_{\Ac\Bc\Cc} = \cF^{(0)}_{\Ac\Bc\Cc} + 3 \delta_{[\Ac \delta} (\htau^\delta)_{\Bc\Cc]}
\end{equation}
at the leading order besides the heterotic flux $\cF^{(0)}_{\Ac\Bc\Cc}$. It is completely (anti-)/chiral and therefore only contributes to the second term of $\Rc_2$ and its conjugate in \eqref{eqn:Rs}. The reason why the second term in \eqref{eqn:THet0} appears is that $\cA$, as the matrix exponent \eqref{eqn:expAp}, contains the identity at the leading order. Hence, one can alternatively write
\begin{equation}
	\Ac = 1 + \Ac^{(>0)}
\end{equation}
because all other terms contain at least one derivative. All non-trivial information about the connection in the \PS{} construction are contained in the second term. We thus define the spin connection
\begin{equation}
	\Omega_{\Ac\Bc\Cc} := A_{\Ac\delta}^{(>0)} (\htau^\delta)_{\Bc\Cc} \,.
\end{equation}

After again eliminating all constant contributions, we are left with the action
\begin{equation}
	S = S_{\cF} + \int \dd x^d e^{-2 \Phi} \Rc_3\,,
\end{equation}
with
\begin{equation}\label{eqn:SF}
	S_{\cF} = \int \dd x^d e^{-2 \Phi} \left( 2 \cD_{\AcR} \cF^{\AcR} - \cF_{\AcR} \cF^{\AcR} + \tfrac12 \cF_{\AcL\BcR\CcR}\cF^{\AcL\BcR\CcR} + \tfrac16 \cF_{\AcR\BcR\CcR} \cF^{\AcR\BcR\CcR} - \text{c.c.}\right)
\end{equation}
and
\begin{equation}
	\Rc_3 = \THet^{\alphaR}{}_{\BcR\CcR} (\htau_{\alphaR})^{\BcR\CcR} + 2 \cD_{\AcR} \Omega_{\BcR}{}^{\BcR\AcR} + \tfrac12 \Omega_{\Ac\BcR\CcR} \Omega^{\Ac\BcR\CcR} + \cF_{\Ac\BcR\CcR} \Omega^{\Ac\BcR\CcR} + 2 \Omega_{\AcR\BcR\Cc} \Omega^{\BcR\AcR\Cc} - \text{c.c.} \simeq 0\,.
\end{equation}
Remarkably, one finds that this term vanishes up to constants by directly computing $\THet^{\alphaR}{}_{\BcR\CcR}$. Therefore, we conclude that the action \eqref{eqn:SF} is invariant -- a result that perfectly matches the expectations set by the gBdR identification. Computing the non-vanishing contributions
\begin{equation}
	\cF^{(0) \alpha\beta\gamma} = - f^{\alpha\beta\gamma}\,, \qquad
	\cF^{(1)}_{ABC} = F_{ABC} \,,
	\qquad \text{and} \qquad
	\cF^{(1)}_{A} = F_A
\end{equation}
at the leading order and combining the last one with
\begin{equation}
	\cD^{(1)}_A = D_A\,,
\end{equation}
we find that the two derivative action matches the expected result from the flux formulation of double field theory, namely
\begin{equation}
	S^{(2)}_{\cF} = \int \dd x^d e^{-2\Phi} \left( 2 D_{\aR} F^{\aR} - F_{\aR} F^{\aR} + \tfrac12 F_{\aL\bR\cR} F^{\aL\bR\cR} + \tfrac16 F_{\aR\bR\cR} F^{\aR\bR\cR} - \text{c.c.} \right)\,.
\end{equation}
At the next order, it is sufficient to just look at the special case $b=0$ due to the $\mathbb{Z}_2$ symmetry highlighted above. The relevant, not yet computed, heterotic fluxes then are
\begin{equation}
	\begin{aligned}
		\cF^{(3)}_{\aR}       & = \frac{a}{4}\left(F^{\bR}F_{\bR}{}^{\cL\dL}F_{\aR\cL\dL}-D^{\bR}(F_{\bR}{}^{\cL\dL}F_{\aR\cL\dL})\right)\, ,                                                          \\
		\cF^{(3)}_{\aR\bR\cR} & = - \frac{3 a}{2}\left(D_{[\aR}F^{\eL\fL}{}_{\bR}-\tfrac{1}{2}F_{\dR[\aR\bR}F^{\dR\eL\fL} - \tfrac{2}{3}F^{\eL}{}_{\dL[\aR}F_{\bR}{}^{\dL\fL}\right)F_{\cR]\eL\fL}\, , \\
		\cF^{(3)}_{\aL\bR\cR} & = -\frac{a}{2}\left(D_{\aL}F^{\eL\fL}{}_{[\bR}+F^{\dR\eL\fL}F_{\aL\dR[\bR}\right)F_{\cR]\eL\fL}\,.
	\end{aligned}
\end{equation}
They need to be combined with
\begin{equation}\label{eqn:cD3}
	\cD^{(3)}_{\aR} = \frac{a}{4} F_{\aR\cL\dL}F^{\bR\cL\dL}D_{\bR}\,,
\end{equation}
and
\begin{equation}
	\cF^{(2)}_{\alphaL\bR\cR} = 2 D_{[\bR} A^{(1)}_{\cR]\alphaL} - A^{(1)}_{\dR\alphaL} F^{\dR}{}_{\bR\cR} - f_{\alphaL}{}^{\betaL\gammaL} A^{(1)}_{\bR\betaL} A^{(1)}_{\cR\gammaL}
\end{equation}
originating from \eqref{eqn:DHet} and \eqref{eqn:F2alpha}, respectively, along with the expansions
\begin{equation}
	\begin{aligned}
		2 \cD_{\AcR}\cF^{\AcR} |^{(4)}                             & =  2 \cD^{(3)}_{\aR}F^{\aR} + 2 D_{\aR} \cF^{(3)\aR}\,,                                               \\
		\cF_{\AcR}\cF^{\AcR} |^{(4)}                               & = 2\cF^{(3)}_{\aR} F^{\aR}\,,                                                                         \\
		\tfrac{1}{6}\cF_{\AcR\BcR\CcR} \cF^{\AcR\BcR\CcR} |^{(4)}  & = \tfrac{1}{3}\cF^{(3)}_{\aR\bR\cR} F^{\aR\bR\cR}\,,                                                  \\
		\tfrac{1}{2} \cF_{\AcL\BcR\CcR} \cF^{\AcL\BcR\CcR} |^{(4)} & = \cF^{(3)}_{\aL\bR\cR}F^{\aL\bR\cR} +\tfrac{1}{2} \cF^{(2)}_{\alphaL\bR\cR}\cF^{(2)\alphaL\bR\cR}\,.
	\end{aligned}
\end{equation}
In this way, we obtain a higher-order contribution $a\mathcal{R}^{-}$ to \eqref{eqn:SF},
where
\begin{equation}\label{eqn:R-}
	\begin{aligned}
		\mathcal{R}^{-}=-\tfrac{1}{2} & \left(D_{\aR}D_{\bR}F^{\bR\cL\dL}F^{\aR}{}_{\cL\dL} + D_{\aR}D_{\bR}F^{\aR\cL\dL}F^{\bR}{}_{\cL\dL} - 2 D_{\aR}F_{\bR}{}^{\cL\dL}F^{\aR}{}_{\cL\dL}F^{\bR}\right.+                                                \\
		                              & \left.  D_{\aR}F^{\aR\cL\dL}D_{\bR}F^{\bR}{}_{\cL\dL}+D_{\aR}F_{\bR\cL\dL}D^{\aR}F^{\bR\cL\dL}-2D_{\aR}F_{\bR}F^{\bR\cL\dL}F^{\aR}{}_{\cL\dL}\right.+                                                             \\
		                              & \left.D_{\aL}F^{\eL\fL}{}_{\bR}F_{\cR\eL\fL}F^{\aL\bR\cR}-D_{\aR}F^{\eL\fL}{}_{\bR}F_{\cR\eL\fL}F^{\aR\bR\cR}- 2 D_{\aR}F^{\aR}{}_{\cL\dL}F_{\bR}{}^{\cL\dL}F^{\bR}\right.-                                       \\
		                              & \left.  4D^{\aR}F^{\cR\bL\dL}F_{\aR\bL}{}^{\eL}F_{\cR\eL\dL}+\tfrac{4}{3}F^{\eL}{}_{\dL\aR}F_{\bR}{}^{\dL\fL}F_{\cR\eL\fL}F^{\aR\bR\cR} +F^{\aR}{}_{\cL\dL}F^{\bR\cL\dL}F_{\aR}F_{\bR}\right.+                    \\
		                              & \left.F_{\aR\dL}{}^{\cL}F_{\bR\cL}{}^{\eL}F^{\aR\dL\fL}F^{\bR}{}_{\fL\eL} - F_{\bR\dL}{}^{\cL}F_{\aR\cL}{}^{\eL}F^{\aR\dL\fL}F^{\bR}{}_{\fL\eL} + F^{\dR\eL\fL}F_{\aL\dR\bR}F_{\cR\eL\fL}F^{\aL\bR\cR}\right)\, ,
	\end{aligned}
\end{equation}
which matches with the literature as can be seen for example by comparing with \cite{Lescano:2021guc,Lescano:2021lup}. As in the last section, the parameter $a$ arises here by collapsing the towers in the last step and adsorbing the resulting $\chiL/g_-^2$. We could continue in the same way with the next order, where this process is more difficult. Because the results will be bulky and not give any further insights, we decided not to perform this computation here.

\section{Conclusions}\label{sec:conclusions}
As demonstrated in the last section, at the end of the day, we recover the results from the original gBdR identification. Still our approach is distinguished from it in different aspects, which we would like to point our here. Perhaps the most important difference is that while the original construction behaves like a black box which only gives the final result, we produce various intermediate results beginning in section~\ref{sec:identification} with the identification. This has two major advantages:
\begin{enumerate}
	\item One can reuse the results from lower orders in the computation of higher orders. Moreover, intermediate results can be checked independently from the final expressions for the action and its gGS transformations.
	\item Expressions in terms of the generalized fluxes on the physical space are, in particular at higher orders, very complicated. Take the gGS transformation in \eqref{eqn:deltaE4final} as an example: In this form, it is nearly impossible to deduce it originates from the much simpler expression \eqref{eqn:gGSmaster} which even suggests a straightforward extension to all orders.
\end{enumerate}
From a practical point of view, they allow to reduce the effort required to compute corrections at higher orders considerably. Furthermore, we obtained new insights in the parameterization of the heterotic frame $\cE$ containing the physical frame $E$ and the connection $\cA$ according to \eqref{eqn:cEtocA}. Different parameterizations are related by field redefinitions. Hence, one can choose them freely. But usually there exists a preferred field basis in which computations simplify. Remarkably, the field basis chosen in \cite{Baron:2020xel} is governed by
\begin{equation}
	\cA = A + \sqrt{1 + A^2}
\end{equation}
and therefore gives rise to the expansion
\begin{equation}
	A' = 1 - \tfrac16 A^3 + \tfrac3{40} A^5 + \dots \,.
\end{equation}
According to \eqref{eqn:Aprime} this would result in $c_1 = - 1/6$, while we instead use $c_1 = 1/3$. We are planning to come back to this point in the future to see if there exists a parameterization which allows to keep the extremely simple form of the gGS transformation \eqref{eqn:gGSmaster} to all orders.

From the point of the \PS{} construction, the choice of $\GS$ imposed by the gBdR identification can be improved by using only the subgroup which keeps the full reduced, twisted torsion covariant. Remember that we lose covariance of the completely (anti-)chiral parts here for the $\GS$ constructed in section~\ref{sec:GS}. In this case, the torsion constraints and the gauge fixing has to be reconsidered whereas the final results should not be affected. But we see that the gBdR identification assumes a symmetry which is more restrictive than it has to be and therefore rules out other deformations as they would be required to obtain $\zeta(3)$ corrections. Hence, we see the results presented here only as the starting point for a deeper exploration of higher-derivative corrections and their relation to dualities. Another important aspect not discussed yet are special geometries, and that particular generalized homogeneous from the backbone of generalized dualities governing certain consistent truncations and most of the known integrable $\sigma$-models. In particular, for the latter, it has been shown that no $\zeta(3)$ corrections arise at four loops due to the special properties of the underlying geometry of the $\eta$- and $\lambda$-deformation of the two-sphere \cite{Alfimov:2021sir}. This might show that, for certain classes of backgrounds, the obstruction described in \cite{Hronek:2020xxi} become irrelevant.

\section*{Acknowledgements}
This work would have been impossible without Daniel Butter, who had the initial idea of recovering the gBdR identification from a twisted version of the \PS{} construction. He not only helped with numerous discussions but also by sharing his notes on the topic. Dan left academia in 2022 and did not had the time to help as much as he wanted in the implementation of this project. He decided that his involvement was not sufficient to be a coauthor of this article. Therefore, we would like to at least honor his contribution here and share the anecdote that internally the adapted version of the \PS{} construction is called the Butter-twist.

Moreover, we would like to thank Eric Lescano for initial collaboration and discussions. Finally, we would like to acknowledge
Martin Cederwall,
Sylvain Lacroix,
David Osten,
Yuho Sakatani,
Luca Scala, and
Linus Wulff,
for discussions directly on the topic or on closely related problems. The work of AG, FH is supported by the SONATA BIS grant 2021/42/E/ST2/00304 from the National Science Centre (NCN), Poland.

\bibliography{literature}

\bibliographystyle{JHEP}

\end{document}